%% file: lbm-gpu-acc.tex
\documentclass[10pt,journal,compsoc]{IEEEtran}

\ifCLASSOPTIONcompsoc
  \usepackage[nocompress]{cite}
\else
  \usepackage{cite}
\fi

\ifCLASSINFOpdf
   \usepackage[pdftex]{graphicx}
   \graphicspath{{figfinal/}{./}}
   \DeclareGraphicsExtensions{.pdf,.jpeg,.png}
\else
   \usepackage[dvips]{graphicx}
\fi

\usepackage[cmex10]{amsmath}
\usepackage{multirow}
\usepackage{algorithm}
\usepackage{algorithmicx}
\usepackage{array}
\usepackage{algpseudocode}
\usepackage{amsmath}

\usepackage{verbatim}

\usepackage{bm}
\usepackage{color}

\begin{document}
%
\title{GPU Optimization for High-Quality\\Kinetic Fluid Simulation}
\markboth{Accepted by IEEE Transactions on Visualization and Computer Graphics}%
{Chen \MakeLowercase{\textit{et al.}}}
%

\author{Yixin Chen,
	Wei Li,
	Rui Fan,
	and Xiaopei Liu
	\IEEEcompsocitemizethanks{
		\IEEEcompsocthanksitem Yixin Chen, Wei Li, Rui Fan and Xiaopei Liu are all with the School of Information Science and Technology (Shanghai Engineering Research Center of Intelligent Vision and Imaging), ShanghaiTech University, China. E-mails: \{chenyx2, liwei, fanrui, liuxp\}@shanghaitech.edu.cn.
}}



\IEEEtitleabstractindextext{%
\begin{abstract}

Fluid simulations are often performed using the incompressible Navier-Stokes equations (INSE), leading to sparse linear systems which are difficult to solve efficiently in parallel.  Recently, kinetic methods based on the adaptive-central-moment multiple-relaxation-time (ACM-MRT) model~\cite{Li-2018,Li-2020} have demonstrated impressive capabilities to simulate both laminar and turbulent flows, with quality matching or surpassing that of state-of-the-art INSE solvers.  Furthermore, due to its local formulation, this method presents the opportunity for highly scalable implementations on parallel systems such as GPUs.
However, an efficient ACM-MRT-based kinetic solver needs to overcome a number of computational challenges, especially when dealing with complex solids inside the fluid domain.  
In this paper, we present multiple novel GPU optimization techniques to efficiently implement high-quality ACM-MRT-based kinetic fluid simulations in domains containing complex solids. Our techniques include a new communication-efficient data layout, a load-balanced immersed-boundary method, a multi-kernel launch method using a simplified formulation of ACM-MRT calculations to enable greater parallelism, and the integration of these techniques into a parametric cost model to enable automated parameter search to achieve optimal execution performance.  We also extended our method to multi-GPU systems to enable large-scale simulations.  To demonstrate the state-of-the-art performance and high visual quality of our solver, we present extensive experimental results and comparisons to other solvers.

\end{abstract}

\begin{IEEEkeywords}
GPU optimization, parallel computing, fluid simulation, lattice Boltzmann method, immersed boundary method
\end{IEEEkeywords}}

\maketitle

\IEEEdisplaynontitleabstractindextext

%
\IEEEpeerreviewmaketitle

\input{1.introduction.tex}
\input{2.related-work.tex}
\input{3.background.tex}

\input{4.our-approach.tex}
\input{5.implementations.tex}
\input{6.results-and-discussion.tex}
\input{7.conclusion.tex}

\section*{Acknowledgments}
We thank all the reviewers for their constructive comments. We also thank Yihui Ma, Chenqi Luo and Chaoyang Lyu from the FLARE Lab at ShanghaiTech University for helping with video preparations. We are grateful to YOKE Intelligence for providing the 3D reconstruction in Fig.~\ref{fig:teaser} for our large scale simulation. This work was supported by the startup fund of ShanghaiTech University and the National Natural Science Foundation of China (Grant No. 61976138).

\ifCLASSOPTIONcaptionsoff
  \newpage
\fi



\bibliographystyle{IEEEtran}
\bibliography{lbm-gpu-acc}

\end{document}

%% file: 1.introduction.tex
\section{Introduction}
\label{sec:intro}


Fluid simulations in computer graphics mostly rely on solving the incompressible Navier-Stokes equations (INSE) using certain types of discretization schemes~\cite{mullen2009energy,zhu2013new,ihmsen2014sph,fu-2017,zehnder-2018}. 
While there are a vast number of available solvers for INSE, including some which achieve relatively high accuracy for turbulent flows (e.g. \cite{zehnder-2018,qu-2019}), most methods require solving large global sparse linear systems, such as the pressure equation or equations resulting from implicit formulations for stability, making these solvers difficult to efficiently parallelize, especially on GPUs or for high-resolution simulations.  For this reason, local but accurate solvers involving no global equations are desirable.

\begin{figure*}[t]
	\centering
	\includegraphics[width=\textwidth]{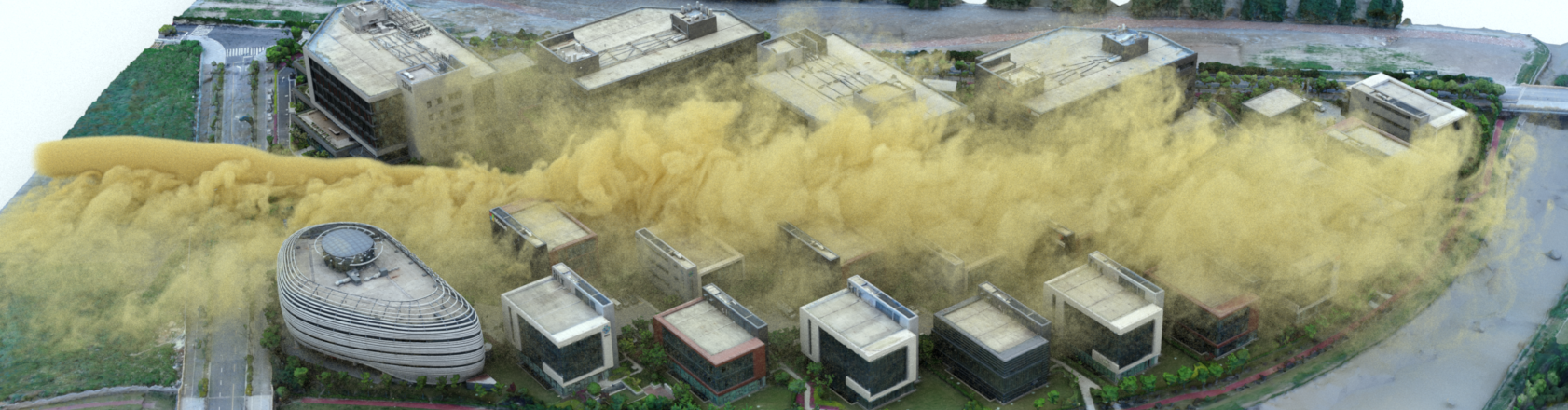} \vspace*{-6mm}
	\caption{\textbf{Simulation of turbulent smoke passing through complex architecture}. With our GPU optimized kinetic solver on a multi-GPU system, we can efficiently simulate this high-resolution scenario (grid resolution 1200$\times$250$\times$840, with 3,688,157 solid samples) at a rate of 1.21 seconds per time step, taking 76,600 time steps in total to produce an animation of 10 seconds.  Here wind moves from left to right through the buildings and smoke is continuously injected from the left and advected by the wind.  Comparably high computational efficiency and visual fidelity is very difficult to achieve using parallel Navier-Stokes solvers at such high grid resolutions.}
	\vspace{-0.2cm}
	\label{fig:teaser}
\end{figure*}

Over the years, kinetic solvers based on the lattice Boltzmann equations (LBE)~\cite{shan-2006} have been considered as an appealing alternative for high-performance fluid simulations. Unlike many INSE solvers, kinetic solvers only rely on local schemes, and thus by nature are well suited to parallelization.  These methods are also conservative by construction, making them theoretically ideal for turbulent flow simulations.  However, despite previous attempts~\cite{Guo_TVCG_2017,wei-2004,Thuerey-2006,Thuerey-2008}, kinetic solvers have seldomly been used in computer graphics applications due to their poor accuracy and stability.  One reason is that most previous works relied on the BGK~\cite{chen1998lattice} or raw-moment multiple-relaxation-time (RM-MRT)~\cite{d2002multiple,liu2012turbulence} models, which are known to exhibit noticeable visual artifacts.  The recent development of the adaptive-central-moment multiple-relaxation-time (ACM-MRT) model~\cite{Li-2018,Li-2020} for kinetic solvers has helped overcome these limitations, leading to an accurate and stable solver demonstrating equal and sometimes superior visual quality compared to state-of-the-art INSE solvers (see Fig.~\ref{fig:teaser}), while also offering the potential for high computational efficiency.
In addition, ACM-MRT-based kinetic solvers can be easily combined with the immersed boundary (IB) method~\cite{Peskin-1972} to enable flexible handling of boundaries with complex geometries for both static and moving solids immersed in the fluid domain~\cite{Li-2020}.

Nevertheless, despite the advancements offered by the ACM-MRT model, a straightforward GPU implementation of it does not produce high performance.
There have been multiple attempts in the literature to improve the parallelism of GPU-based kinetic solvers, mostly relying on optimizations to data layout~\cite{herschlag2018gpu,calore2019optimization} or the order of computations~\cite{tolke2008teraflop}.  However, these works targeted the traditional BGK~\cite{Chen-1998} and RM-MRT models~\cite{d2002multiple}, which typically cannot attain the visual quality needed for graphical fluid simulations.  Furthermore, to the best of our knowledge, there is no work on GPU optimizations for enhancing the performance of IB-based kinetic solvers to support the immersion of complex dynamic solids in fluids.


In this paper, we aim to increase the performance of kinetic solvers using the ACM-MRT model and IB method on GPUs by addressing several key issues.  First, when using the IB method in a kinetic solver, the data layout should be designed to maximize the correspondence between spatial locality in the simulation domain and locality in GPU memory.  For this purpose, we propose a new parametric data layout model which seeks to maximize data access contiguity.  Second, a straightforward parallel implementation of the IB method may introduce substantial load imbalance due to large variations in the number of solid samples around different fluid nodes.  We propose a new, more efficient method to implement IB which parallelizes over solid samples instead of fluid nodes to significantly improve load balance.
Finally, the ACM-MRT model performs lengthy calculations which require a large number of registers for each GPU thread, leading to reduced occupancy and execution efficiency.  We perform multiple kernel launches and simplify the ACM-MRT mathematical expressions to mitigate these issues.  These multifaceted optimizations are integrated into a single parametric cost model, allowing us to perform an automated search in the parameter space to obtain the best execution performance.  We also extend our techniques to multi-GPU systems and show that our solver can easily and efficiently support high-resolution large-scale simulations.

In summary, we make the following key technical contributions to high-performance fluid simulations using IB-based kinetic solvers with the ACM-MRT model:
\vspace{-0.1cm}
\begin{itemize}
	\item A new parametric data layout to improve memory performance for IB-based kinetic solvers, improving access contiguity for both fluid and solid data.	
	\item An optimized parallel IB implementation with separate domain boundary treatments to improve load balancing and reduce warp divergence.
	\item Tuning the number of kernel launches and simplifying the mathematical formulation of the ACM-MRT model to improve GPU occupancy and execution efficiency.
	\item An integrated cost model with automated parameter search to produce the best settings for each simulation instance.
\end{itemize}

With the above optimizations, our new implementation of a kinetic solver for the ACM-MRT model can run 5 to 10 times faster than a baseline GPU implementation.  It also significantly outperforms  state-of-the-art GPU-based INSE solvers, being about an order of magnitude faster at the same grid resolution and similar visual quality.  Due to its faster rate of convergence, our solver can even produce visual quality similar to INSE solvers while using a significantly lower grid resolution, which leads to a two order of magnitude improvement in performance.  High-resolution fluid simulations on multi-GPU systems can also be efficiently implemented.  Fig.~\ref{fig:teaser} shows an example of a large-scale multi-GPU simulation at a resolution of $1200\!\times\!250\!\times\!840$.   We also present a number of performance comparisons, analyses and visual animations to demonstrate the capabilities of our solver to produce high quality, robust fluid simulations at practically useful speeds.  


%% file: 2.related-work.tex
\section{Related Work}
\label{sec:related_work}
Our work is related to high-performance fluid simulations on the GPU.
We first review different fluid simulation methods for both INSE and kinetic solvers, and then review different optimization and parallelization techniques.

\vspace{-0.1cm}
\subsection{Fluid Simulations}
\subsubsection{Incompressible Navier-Stokes solvers}
There has been a vast number of INSE solvers developed in the computer graphics community.
Early works stemmed from Stam~\cite{stam1999stable} and used stable semi-Lagrangian advection, but suffered from excessive numerical diffusion.
This issue was progressively improved using higher-order schemes~\cite{kim2005flowfixer,wang2008observations,selle2008unconditionally,cui-2018,zehnder-2018,qu-2019}.
Another way to counteract numerical diffusion is to use vortex-based methods~\cite{fedkiw2001visual,park2005vortex,weissmann2010filament,pfaff2010scalable,brochu2012linear,golas2012large,zhang2014pppm,zhang-2015}, where the velocity field is corrected by vorticity distributions.
Noise-based methods~\cite{bridson2007curl,kim2008wavelet} are a cheap way to preserve flow details, but the resulting simulation may be non-physical and appear unrealistic.  These approaches have recently been improved using neural network-based methods~\cite{jeong2015data,chu2017data,xie-2018,Bai-2020}.
While the above works employed only uniform grids, non-uniform grids~\cite{losasso2004simulating,zhu2013new,setaluri2014spgrid} or even tetrahedral meshes~\cite{mullen2009energy,clausen-2013,ando2013highly} have been used to perform adaptive computations.

In addition to grids, INSE can be solved using particle methods~\cite{becker2007weakly,solenthaler2009predictive,akinci2012versatile,ihmsen2014sph,winchenbach2017infinite}, with more recent techniques allowing for volume conservation~\cite{de2015power}, improved incompressibility~\cite{bender-2016} and better boundary treatments~\cite{band-2017,Band-2018}.
To retain the benefits of both grids and particles for better advection and incompressibility, hybrid methods combining particles and grids have been developed~\cite{jiang2015affine,zhang2014pppm,zhang2016resolving,zhu2005animating,fu-2017}.
Note that particle-based methods without an auxiliary grid may not need global sparse linear system solvers, but they usually have difficulty simulating turbulent flows unless a very large number of particles is used.


\subsubsection{Kinetic solvers} 
Instead of solving INSE directly, kinetic methods solve the more general Boltzmann equation, which approximates INSE with different degrees of accuracy depending on how relaxations in collisions is modeled.  Early kinetic solvers used the lattice BGK (or single-relaxation-time) model~\cite{chen1998lattice}.  To improve accuracy and stability, raw-moment multiple-relaxation-time (RM-MRT) model~\cite{d2002multiple} was developed, but it had problems with turbulent flows due to violation of Galilean invariance, and leads to strong dispersion errors at high Reynolds numbers.
Recent improvements in relaxation modeling include the cascaded relaxation~\cite{geier2006cascaded,lycett2014binary} and central-moment multiple-relaxation-time (CM-MRT) models~\cite{geier-2009,shan-2019}, especially the non-orthogonal version~\cite{de2017nonorthogonal}, which much better respects Galilean invariance.
However, these works still failed to address how to determine high-order relaxation parameters, which has been found important for turbulent flow simulations.
Based on observations and analyses of the non-orthogonal CM-MRT model, Li et al.~\cite{Li-2018,Li-2020} proposed adaptive schemes for relaxation parameters based on local flow characteristics.  This model, called ACM-MRT, allows for much more accurate simulations of both laminar and turbulent flows. 


When solids are immersed inside the fluid domain, boundary conditions need to be satisfied.  Bounce-back schemes have been used to enforce no-slip~\cite{rohde-2003} or slipping~\cite{sbragaglia-2005} conditions.
To improve accuracy, curved boundary conditions were developed~\cite{verschaeve-2010}, which can also be extended to support dynamic solids~\cite{mei-1999}.
However, these methods may encounter serious problems for turbulent flows when grid resolution around the solid boundary is not sufficiently high.
A more flexible technique for boundary treatment is based on the immersed boundary method~\cite{Peskin-1972,feng-2004,wu-2010}, which can be used to handle both static and dynamic solid boundaries using penalty forces.


\subsection{Parallel Performance Optimizations}

\subsubsection{Parallel Navier-Stokes solvers}
As mentioned earlier, most INSE solvers require solving large global equations and are therefore difficult to parallelize.  
Parallel INSE solvers for multicore systems were developed using OpenMP~\cite{sato-2013} and extended to multi-node cluster systems using MPI~\cite{guo-2015}.
In computer graphics, Mashayekhi et al.~\cite{mashayekhi-2018} proposed a system called ``Nimbus'', which automatically distributes grid-based and hybrid simulations across cloud computing nodes for faster execution at higher grid or particle resolutions.

INSE solvers can also be parallelized on the GPU.
Early works tried to map the entire computation into texture memory and use fragment programs to solve INSE~\cite{brandvik-2008,liu-2004,kruger-2003}.
With the advent of CUDA~\cite{Cuda}, optimized parallel sparse linear solvers became available~\cite{cusparse} on single and multi-GPU systems, making GPU implementations of many INSE solvers much simpler~\cite{thibault-2009,griebel-2010,henniger-2010,Mcadams-2010}.

Recently, GPU optimizations for the parallel material-point method (MPM) were proposed to accelerate a hybrid fluid solver~\cite{gao-2018}.
In addition, an efficient large-scale fluid simulator on the GPU using the fluid‐implicit
particle (FLIP) method~\cite{wu-2018} and a high-level, data-oriented programming language for sparse data structures (Taichi)~\cite{hu-2019} has been proposed.
To utilize the computational advantages of both CPUs and GPUs, some works proposed parallel INSE solvers on hybrid heterogeneous systems~\cite{alfonsi-2011,wang-2014,posey-2013}.


\subsubsection{GPU optimizations for parallel kinetic solvers}
While INSE solvers are inherently difficult to parallelize, kinetic solvers use a local formulation and thus are naturally scalable and well suited for implementation on highly parallel GPUs. 
Similar to earlier parallelization methods for INSE solvers on the GPU, texture and render buffers were initially used to accelerate kinetic solvers~\cite{li2003implementing,zhao2007flow}.
Using CUDA, GPU implementations of kinetic solvers became much simpler.  
However, a straightforward implementation of a kinetic solver does not achieve very good performance.
One issue is memory coalescing, which is crucial for high GPU performance.  The array-of-structures (AoS) data layout, which performs well on CPUs~\cite{WelleinOn2006}, was replaced by a structure-of-arrays (SoA) data layout on the GPU~\cite{tolke2008teraflop}. 
The collected SoA (CSoA) data layout, which combines AoS and SoA, was recently proposed~\cite{calore2019optimization} to further enhance performance and allow better caching when solid boundaries are complex or domain boundaries are irregular~\cite{herschlag2018gpu}.

To reduce non-coalesced memory accesses, the use of a single kernel launch with swapping for both the streaming and collision stages of a simulation has been proposed~\cite{MattilaAn2007}. Since non-coalesced reads require less time than non-coalesced writes, the pull-in scheme was argued to be better than the push-out scheme~\cite{delbosc2014optimized}.
With the above optimizations, kinetic solvers using different lattice models (D2Q9~\cite{tolke2010implementation}, D3Q13~\cite{tolke2008teraflop}, D3Q19~\cite{bailey2009accelerating,habich2011performance}) were proposed on the GPU, based mostly on the BGK or RM-MRT collision models. 
In addition, to simulate larger scenarios, multi-GPU kinetic solvers were proposed using CUDA~\cite{bernaschi2010flexible, obrecht2013multi} and OpenMP~\cite{myre2011performance}.
To our knowledge, there has been no work on GPU optimizations for kinetic solvers with the D3Q27 lattice for either the BGK or RM-MRT collision models.

\vspace{0.2cm}
\textit{Our contributions.}
Most of the GPU optimizations for kinetic solvers described above were for fluids only.  When complex solids are present in fluids and boundary conditions are enforced using the immersed boundary method, the  previous techniques may lead to poor performance due to sub-optimal data layout, load imbalance and warp divergence.  Furthermore, when the ACM-MRT model is used to achieve high visual quality, a straightforward implementation may result in low GPU occupancy and further depress performance.
This paper addresses these issues and presents a simulator which is substantially faster than state-of-the-art GPU-based INSE solvers, while at the same time producing equal or higher visual quality.

%% file: 3.background.tex
\section{Computational Framework}
\label{sec:background}
\begin{figure}[t]
	\centering
	\includegraphics[width=0.6\columnwidth]{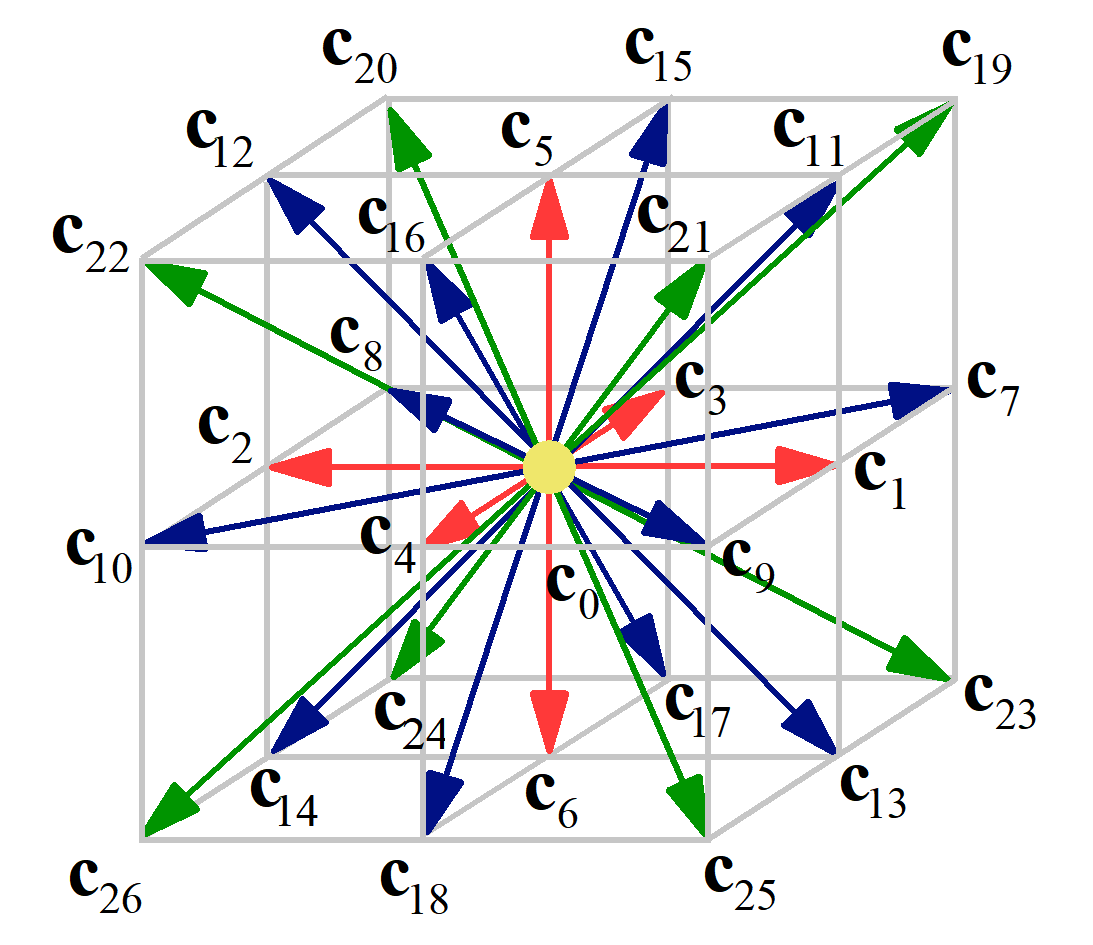}
	\vspace{-4mm}
	\caption{\textbf{Lattice structure.} D3Q27 lattice structure, where $\bm{c}_i$'s are discretized microscopic velocities associated with the corresponding $f_i$.}\vspace{-2mm}
	\label{fig:lattice_structure}
\end{figure}

Before introducing our GPU optimizations in detail, we first briefly review the computational framework employed for fluid simulations in this paper.
%
Fluid flows are usually modeled using the Navier-Stokes (NS) equations, but can alternatively be described  by the Boltzmann equation (BE)~\cite{harris-2004},
which is more general and can recover the NS equations under certain constraints~\cite{shan-2006}.
BE can usually be solved by a set of lattice Boltzmann equations (LBE)~\cite{shan-2006}, which are local and conservative with respect to density and momentum:
\begin{equation}\label{eq:lbm}
f_i(\bm{x}_k + \bm{c}_i,t + 1) - f_i(\bm{x}_k, t) = \Omega_i(\rho, \bm{u}) + G_i(\bm{g}).
\end{equation}
Here, $k$ is the position index of each grid node; $t$ $\in$ $\{1,2,...\}$ is the time step index, and $i$ $\in$ $\{0,1,...,26\}$ is the index of discrete particle velocities $\bm{c}_i$ used in the D3Q27 lattice structure\footnote{See Fig.~\ref{fig:lattice_structure} for an illustration, and the \textit{supplementary document} for the specific values of $\bm{c}_i$.} of the ACM-MRT model.  Each $\bm{c}_i$ is associated with a corresponding discretized distribution function $f_i$, and $\bm{x}_k\!+\!\bm{c}_i$ is exactly located at a nearby node.
Finally, $\Omega_i$ and $G_i$ are the discretized collision and external force terms, and the macroscopic density $\rho$ and velocity $\bm{u}$ can be obtained by taking the following moments w.r.t $\bm{c}_i$:
\begin{equation}\label{eq:discrete_lbm_moments}
\rho = \sum_i f_i ,\;\;\;\;\; \bm{u} = \frac{1}{\rho}\sum_i \bm{c}_if_i .
\end{equation}

Due to its simple algebraic formulation, the solution procedure for LBE can usually be split into the following three steps, starting with \emph{streaming}:
\begin{equation}
f_i^*(\bm{x},t) = f_i(\bm{x}-\bm{c}_i, t).
\label{eq:lbm_streaming}
\end{equation}
Then, macroscopic quantities ($\rho$ and $\bm{u}$) are computed (Eq.~\ref{eq:discrete_lbm_moments}) and boundary treatment is applied.
Finally, we perform the \emph{collision} step with an external force term $G_i$:
\begin{equation}
f_i(\bm{x},t+1) = f_i^*(\bm{x},t) + \Omega_i(\rho^*, \bm{u}^*) + G_i,
\label{eq:lbm_collision}
\end{equation}
where $\rho^*$ and $\bm{u}^*$ are computed by the streamed distribution functions $f_i^*$ after applying boundary treatments.

Key to ensuring accurate and stable fluid simulations for kinetic solvers is the formulation of $\Omega_i$.  Here, we employ the ACM-MRT model~\cite{Li-2018,Li-2020}, which is constructed by a relaxation process with multiple rates:
\begin{equation}
\bm{\Omega} = -\bm{M}^{-1}\bm{D}\bm{M}(\bm{f}- \bm{f}^{eq}) = -\bm{M}^{-1}\bm{D}(\bm{m}- \bm{m}^{eq}),
\label{eq:lbm-mrt-relaxation-moments}
\end{equation}
where for each grid node, $\bm{f}$, $\bm{f}^{eq}$,  $\bm{m}$, $\bm{m}^{eq}$ and $\bm{\Omega}$ are vectors containing the assembly of distribution functions $f_i$, their equilibrium $f_i^{eq}$, their corresponding distributions in moment space $m_i$ and $m_i^{eq}$, and the collision results $\Omega_i$, respectively. 
$\bm{M}$ is a 27$\times$27 moment-space projection matrix, and $\bm{D}$ is a diagonal matrix with the same dimensions containing different relaxation rates.
Low-order relaxation rates are determined by physical parameters (e.g., kinematic viscosity $\nu$), while high-order relaxation rates should be determined adaptively, balancing local dissipation and dispersion errors (see \cite{Li-2018} and \textit{supplementary document} in \cite{Li-2020}).



\begin{figure}[t]
	\centering
	\includegraphics[width=0.9\columnwidth]{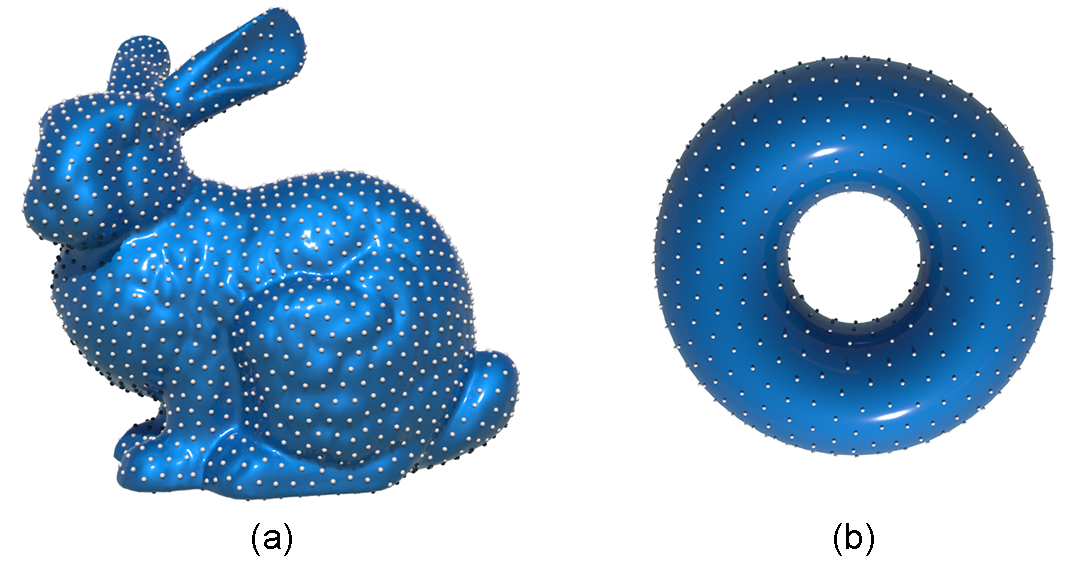}
	\vspace{-4mm}
	\caption{\textbf{Surface sampling used for the immersed boundary method.} We employ Poisson-disk sampling to generate uniformly distributed solid samples over the surface, which are used to enforce boundary conditions at the solid boundary.  Note that the samples here are for illustration only; in practice the sampling should be much denser.}\vspace{-2mm}
	\label{fig:sample}
\end{figure}

\subsection{Boundary treatment}
\label{sec:boundarytreat}

\begin{figure}[t]
	\centering
	\includegraphics[width=0.85\columnwidth]{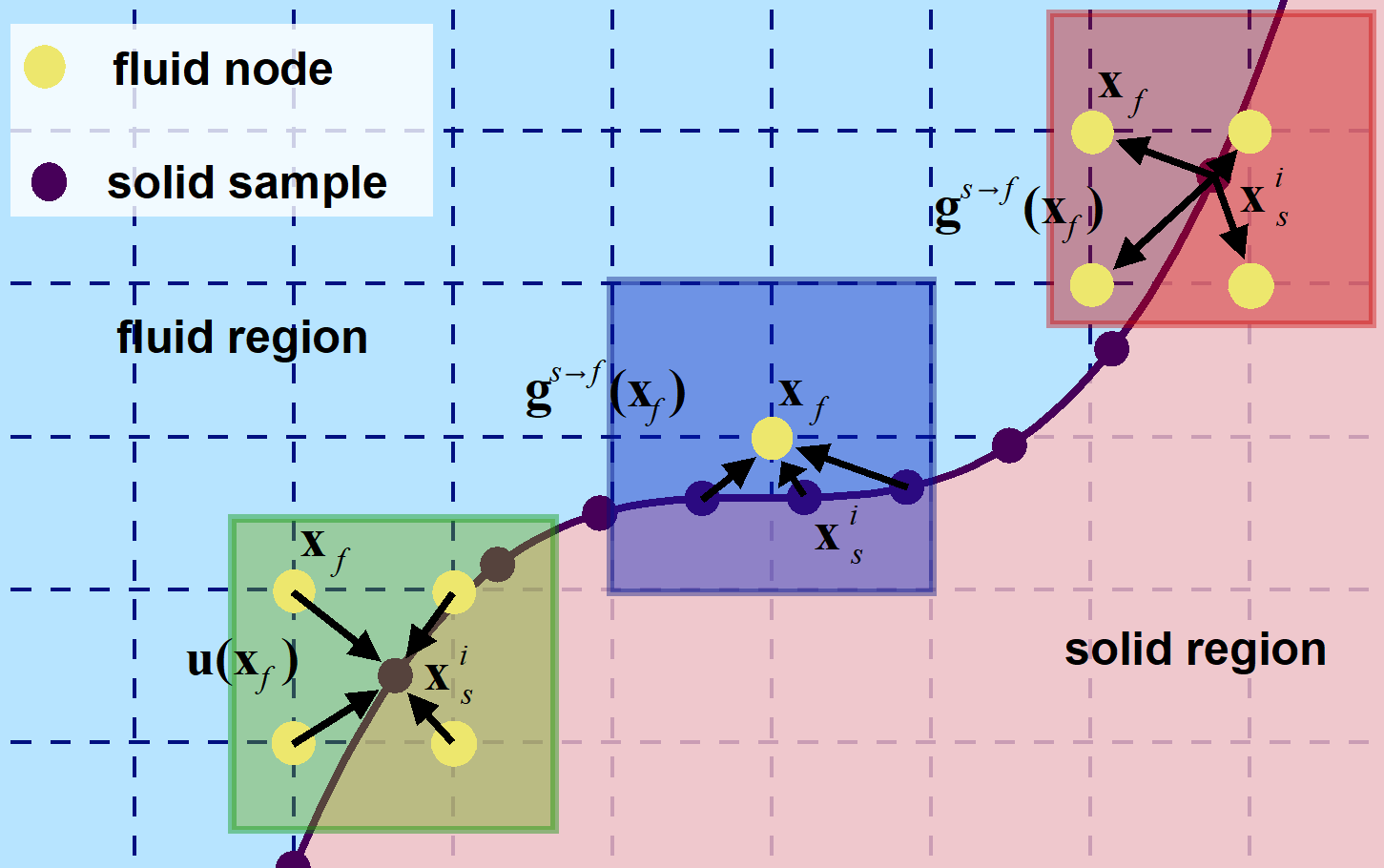}
	\vspace{-2mm}
	\caption{\textbf{Implementing the immersed boundary method.}  We illustrate two ways to compute the penalty force at each fluid node bordering the solid boundary. The fluid velocity is first interpolated from the neighboring fluid nodes (green box). Then to compute the penalty force at each fluid node: 1) boundary samples are first found in a region around the fluid node, possibly using a grid accelerated data structure, then the forces are summed with a spreading kernel to form the force at the fluid node (blue box); or 2) starting from boundary samples, their penalty forces are added directly to the nearby fluid nodes (red box).}
	\label{fig:ib-method}
\end{figure}

When solids are immersed inside fluids, boundary conditions need to be satisfied.
The traditional method for boundary treatment in kinetic solvers is to use bounce-back schemes~\cite{girimaji2012lattice,Thuerey-2006}.
However, when dealing with curved or moving solids, these schemes are either inaccurate or inflexible.  An alternative method which balances flexibility and accuracy for boundary treatment is the immersed boundary (IB) method~\cite{Peskin-1972,feng-2004,wu-2010, Li-2020} we employ in this paper, which usually follows three main steps:
\begin{itemize}
	\item{\textit{Solid surface sampling}.}
	First, we uniformly sample the solid surface using Poisson-disk surface sampling~\cite{yuksel-2015}, see Fig.~\ref{fig:sample}.
	The sampling should be dense enough such that each fluid cell around the solid boundary contains 10 to 100 samples (depending on the local geometry of the solid) to ensure sufficient accuracy.
	\item{\textit{Force computation at solid samples}.}
	A solid sample $\bm{x}_s$ may not be located exactly at a fluid grid node (see Fig.~\ref{fig:ib-method} for a 2D illustration).  Thus we first interpolate fluid velocities $\bm{u}(\bm{x}_s)$ from nearby fluid nodes (cf. the green box in Fig.~\ref{fig:ib-method}) using a kernel function $K(\cdot)$:
	\begin{equation}\label{eq:IB-LBM-interpolation}
	\bm{u}(\bm{x}_s) = \sum_{i \in N(\bm{x}_s)} K(\|\bm{x}_s - \bm{x}_f^i\|) \bm{u}(\bm{x}_f^i).
	\end{equation}
	$\bm{u}(\bm{x}_s)$ may be different from the desired boundary velocity $\bm{u}_b (\bm{x}_s)$, so a penalty force should be applied to the fluid at the solid sample location $\bm{x}_s$:
	\begin{equation}\label{eq:IB-LBM solid-to-fluid}
	\bm{g}^{s \rightarrow f}(\bm{x}_s) = \rho [\bm{u}_b(\bm{x}_s)-\bm{u}(\bm{x}_s)].
	\end{equation}
	\item{\textit{Force spreading to fluid nodes}.}
	Since $\bm{g}^{s \rightarrow f}(\bm{x}_s)$ is applied at the solid sample, we need to spread it to the nearby fluid nodes $\bm{x}_f$ using the same kernel:
	\begin{equation}\label{eq:IB-LBM-spreading}
	\bm{g}^{s \rightarrow f}(\bm{x}_f) = \sum_{i \in N(\bm{x}_f)} K(\|\bm{x}_s^i - \bm{x}_f\|) \bm{g}^{s \rightarrow f}(\bm{x}_s^i).
	\end{equation}
	We use the smallest 2$\times$2$\times$2 kernel in 3D (and 2$\times$2 kernel in 2D) to preserve vortices around solids~\cite{kruger-2011}.
\end{itemize}
Note that the number of solid samples involved in the spreading process depends on the locations of the samples, and can vary significantly over the solid surface. In addition, with the IB method, solids are allowed to move inside fluids, creating turbulent flows at high speeds or with low viscosity.  See the supplementary document for more details of the IB method used in the kinetic solver.

\subsection{A Baseline GPU Implementation}
\label{sec:straight_implementation}
With the procedures described above, we can produce a non-optimized baseline GPU implementation of an IB-based kinetic solver with the ACM-MRT model as follows:
\begin{itemize}
	\item \textit{Initialization.} Before starting the simulation, we allocate global memory on the GPU for the $f_i$ values at times $t$ and $t+1$, and for the density $\rho$, velocity $\bm{u}$ and external force $\bm{g}$ at time $t$, for each fluid node in the simulation.  Vector fields (e.g., $\bm{f}$ and $\bm{u}$) are stored as linear arrays of vectors.
	\item \textit{Streaming.} At each simulation time step $t$, we parallelize over the fluid nodes, and advect $f_i$ from the neighbors of each fluid node (cf. Eq.~\eqref{eq:lbm_streaming}) by reading and writing to GPU global memory.
	
	\item \textit{Calculating macroscopic fields.} After streaming, we again parallelize over the fluid nodes and compute the density $\rho$ and velocity $\bm{u}$ at each node (cf. Eq.~\eqref{eq:discrete_lbm_moments}).
		
	\item \textit{Boundary treatment.}
	We employ the IB method for boundary treatment.  When computing penalty forces at solid samples (cf. Eqs.~\eqref{eq:IB-LBM-interpolation} and \eqref{eq:IB-LBM solid-to-fluid}), we parallelize over the sample points and access the nearby fluid nodes (cf. the green box in Fig.~\ref{fig:ib-method}).  However, when spreading forces (cf. Eq.~\eqref{eq:IB-LBM-spreading}), we parallelize over the fluid nodes in the bounding box(es) of the solid object(s) (cf. the blue box in Fig.~\ref{fig:ib-method}) and access the nearby solid samples. Note that this parallelization procedure directly follows the order of the equations defining the IB method.
	
	\item \textit{Collision.} After boundary treatment, we perform collisions using $\Omega_i$ and the force term $G_i$ (cf. Eq.~\eqref{eq:lbm_collision}), which is done by parallelizing over the fluid nodes.  We perform all collision computations in a single kernel, in keeping with previous GPU-based lattice Boltzmann implementations.
	
	\item \textit{Update.} Finally, with $\Omega_i$ and $G_i$ computed, we update $f_i$ and move to time step $t+1$.
	
\end{itemize}
Lastly, note that the matrix-based computations during the collision step (cf. Eq.~\ref{eq:lbm-mrt-relaxation-moments}) are typically implemented by fully expanding the computation into arithmetic operations between scalar values, rather than through calls to matrix multiplication subroutines.

%% file: 4.our-approach.tex
\section{High-Performance GPU Optimizations}
\label{sec:method}
The baseline GPU implementation described in Sec. \ref{sec:straight_implementation} is expected to be highly scalable due to the local formulation of the kinetic solver.
However, as we discussed in Sec.~\ref{sec:intro}, a number of optimizations can be performed to further improve its speed by more than an order of magnitude in certain simulation scenarios.  In this section, we present a number of optimization techniques, and also show how to integrate these techniques into an automated optimal parameter searching framework to enable fast fluid simulations.


\subsection{GPU Thread Assignment}
Before introducing our GPU optimizations in detail, we first describe how threads are usually assigned in GPU-based kinetic fluid simulations containing immersed solids.  Depending on which part of the simulation is being performed, a GPU thread can be assigned to perform the work at either a fluid node or a solid sample.  As fluid nodes are typically laid out in a 3D grid, we use 3D thread blocks to index threads assigned to the fluid nodes.  We ensure that the size of the first dimension of the thread block is a multiple of the GPU warp size (typically 32), while the sizes of the other dimensions are user tunable and depend on the type of GPU being used.  Solid samples, while used to describe a 3D surface, are typically specified using a 1D index.  Therefore, we also use 1D indices, possibly with a permuted ordering, for threads assigned to solid samples.

\subsection{Improving Memory Performance}

A number of previous works on GPU optimized kinetic solvers~\cite{mattila2008comparison, valero2015accelerating,herschlag2018gpu} have pointed out the importance of data layout on simulation performance.  GPU main memory is laid out in \emph{segments} of consecutive addresses, typically 128 bytes in size.  GPU threads execute in units called \emph{warps}, and a warp of threads accessing data from a single memory segment achieves much higher performance than the warp accessing data from multiple segments.  Thus, an important principle in GPU memory optimization is \emph{memory coalescing}, i.e., ensuring a warp of threads access contiguous, segment-aligned memory locations.

\subsubsection{Optimal data layout for fluid nodes}
\begin{figure}[t]
	\centering
	\includegraphics[width=\columnwidth]{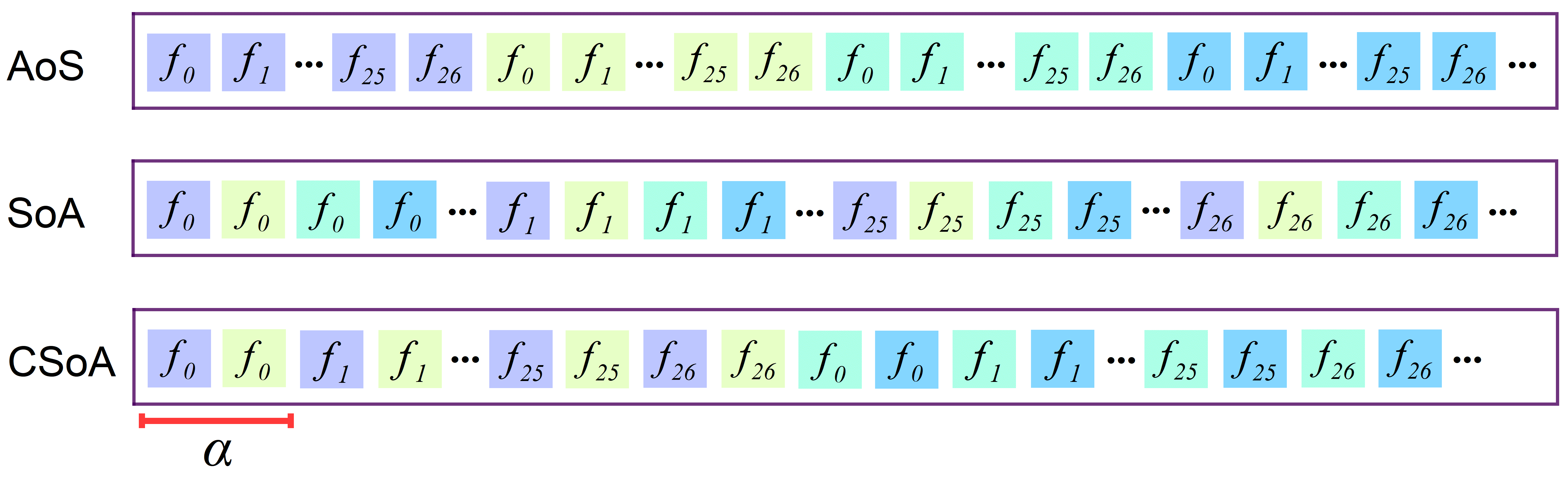}
	\vspace{-8mm}
	\caption{\textbf{Different data layouts of fluid nodes in memory.} AoS: All $f_i$ values from a node are stored consecutively, followed by $f_i$ values of subsequent nodes. SoA: For each $i \in \{0, \ldots, 26\}$, $f_i$ values for all the nodes are stored consecutively. CSoA: For each $i \in \{0, \ldots, 26\}$, the $f_i$ values of $\alpha$ nodes, for some choice of $\alpha$, are stored consecutively.}
	\label{fig:AoSoA}
\end{figure}

We first look at how data for fluid nodes should be laid out in GPU memory.  Fluid data can be described using a number of vector fields, e.g., for different $f_i$'s, $\bm{u}$ and $\bm{g}$.  Collections of vectors are usually stored in one of the two ways, as an array-of-structures (AoS)~\cite{WelleinOn2006}, where each element in an array is a structure storing all vector components of a fluid node (cf. Fig.~\ref{fig:AoSoA} (top) for an AoS layout of the $f_i$ vectors), or as a structure-of-arrays (SoA), where each vector component of the fluid nodes over the entire domain is first grouped together and then stored consecutively in an array (cf. Fig.~\ref{fig:AoSoA} (middle)).  Recall that we assign one thread to process each fluid node.  During the streaming and collision steps of the simulation, threads need to access $f_i$ values from neighboring nodes.  To coalesce these memory accesses, many works adopt the SoA layout for fluid data~\cite{tolke2008teraflop,bernaschi2010flexible}.  Recent works have proposed the CSoA data layout~\cite{herschlag2018gpu,calore2019optimization} to further improve coalescing and caching effects.  Let $\alpha$ denote the number of fluid nodes that we collect into a group (cf. Fig. 5 (bottom)) and $\beta$ the number of components that each fluid node vector contains (e.g., $\beta = 27$ for $\bm{f}$ in the D3Q27 model).  Then the $i$'th component of the vector for the $k$'th fluid node is stored at the following memory location:
\begin{equation}
\beta \alpha \lfloor \frac{k}{\alpha} \rfloor + \alpha i + k \! \mod \alpha .
\label{eq:indexing}
\end{equation}

\subsubsection{Optimal layout for the immersed boundary method}
\label{sec:optimal_ib_data_layout}

\begin{figure}[t]
	\centering
	\includegraphics[width=0.97\columnwidth]{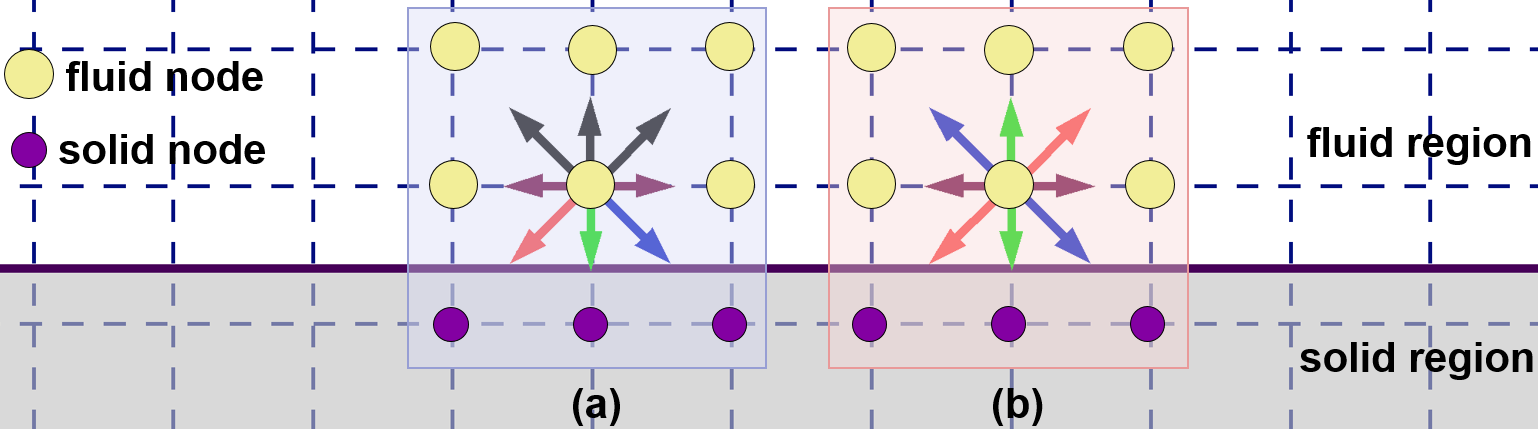}
	\vspace{-2mm}
	\caption{\textbf{Bounce-back scheme for a non-slip boundary.} During streaming at each fluid node, some nearby grid nodes may be located inside the solid region (e.g. the purple nodes in (a) and (b)), and the distribution functions (gray arrows) are unknown and need to be reconstructed.  The bounce-back scheme uses the corresponding distribution functions in the opposite directions (arrows with the same color in (b)) to fill in these unknown distribution functions, ensuring the no-slip condition. Since the solid boundary location around a fluid node may be arbitrary, conditional statements are needed in the kernel code to identify which set of nearby grid nodes are inside solid regions and require bounce-back treatment. Since different kernel threads may follow different conditional branches, warp divergence may occur to slow down the simulation.}
	\label{fig:bounce_back_method}
\end{figure}

When solid objects are immersed inside the fluid domain, boundary conditions need to be satisfied at the solid surface.
As stated in Sec.~\ref{sec:intro}, we use the IB method to enforce boundary conditions instead of the traditional bounce-back scheme.
This is because the bounce-back scheme (as shown in Fig.~\ref{fig:bounce_back_method} and Alg.~\ref{alg:Bounce-back-method}) requires each grid location to be labeled as either solid or fluid, leading to conditional identification of the distribution functions and warp divergence, which slows down the simulation.  The IB method on the other hand only uses penalty forces around solid boundaries, which allows overlap of solid and fluid regions and does not require conditional identification of solid nodes, leading to better parallel performance.  

\begin{algorithm}[t]
	\small
	\caption{Bounce-back scheme in kinetic solver}\label{alg:Bounce-back-method} 
	\begin{algorithmic}[1]
		\If {a grid node at $\bm{x}$ is fluid node}
		\For {each direction $\bm{c}_i$ of the node}
		\If {the grid node at $\bm{x}-\bm{c}_i$ is solid node}
		\State set $f^*_i(\bm{x},t)$ = $f_{i'}^*(\bm{x},t)$;	\Comment{$f^*_{i'}$ is opposite to $f^*_i$.}
		\EndIf
		\EndFor
		\EndIf
		\Comment{Bounce-back scheme on $f^*_i$ after streaming.}
	\end{algorithmic}
	\label{alg:bounce-back-scheme}
	\vspace{0.1cm}
\end{algorithm}

The main problem we encountered when dealing with solids is that solid samples may initially be stored in an arbitrary order in GPU memory, depending on how the solid surface was triangulated or how samples within each triangle were ordered.  Recall from Sec.~\ref{sec:background} that velocity is interpolated at each solid sample from nearby fluid nodes and penalty forces are spread to fluid nodes from solid samples.  Computing these quantities requires solid samples to access nearby fluid data, and hence the order of the solid samples affects the memory access pattern and performance.  Note that this behavior has not been previously discussed in the literature.

\begin{figure}[t]
	\centering
	\includegraphics[width=\columnwidth]{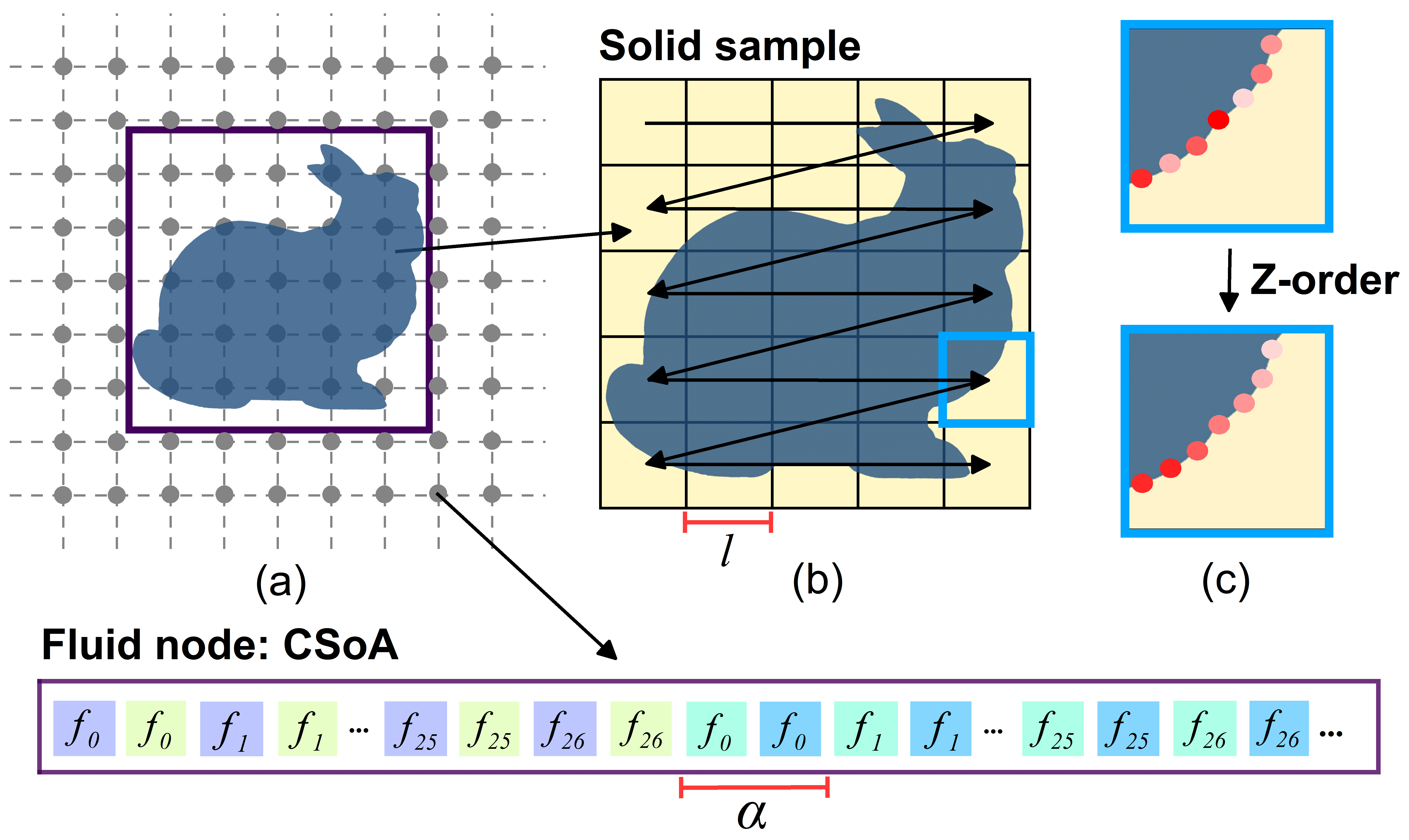}
	\vspace{-4mm}
	\caption{\textbf{2D illustration of the data layout for fluid nodes and solid samples in memory.} To store solid samples, we (a) first find the bounding box of the solid, and then (b) divide the box into blocks of size $\ell\!\times\!\ell$ (in 3D we use $\ell\!\times\!\ell\!\times\!\ell$). These blocks are stored in the same order as the fluid nodes, as shown by the zig-zag black arrows. Within each block, solid samples are first stored in the order they are sampled based on the original triangulation. To further improve memory access performance, these solid samples are Morton-coded and stored in Z-order, where the shading of each sample in (c) represents the ordering of the samples in memory (lighter colors indicate smaller indices). The fluid nodes are stored in the CSoA format, with interpolation parameter $\alpha$ specifying the number of repeating $f_i$'s, i.e. the number of fluid nodes in each group.}
	\label{fig:ib-lbm-layout}
\end{figure}

To alleviate the above problem, we observe that by processing nearby solid samples using consecutive threads, it may be possible to improve cache hit rates for fluid data accesses and improve memory performance.  Based on this idea, we define a parameter $\ell$ and partition the fluid grid around the solid into \emph{blocks} of size $\ell\!\times\!\ell\!\times\!\ell$, ordering these blocks by the natural $x\!-\!y\!-\!z$ order.  We then order the solid samples based on the block they lie in, so that samples from an earlier block are stored before samples from a later block.  Within each block, we order the samples using Z-ordering, by computing the Morton code~\cite{morton-1966} for each sample and sorting the codes within the block (cf. Fig.~\ref{fig:ib-lbm-layout}(c)).  We determine the optimal $\ell$ using our parametric cost model, as described later in Sec.~\ref{sec:parametric_model}.
Note that this is a new data layout for improving coalescing, and has not been proposed in previous IB-based implementations of the kinetic method.

\begin{figure}[t]
	\centering
	\includegraphics[width=\columnwidth]{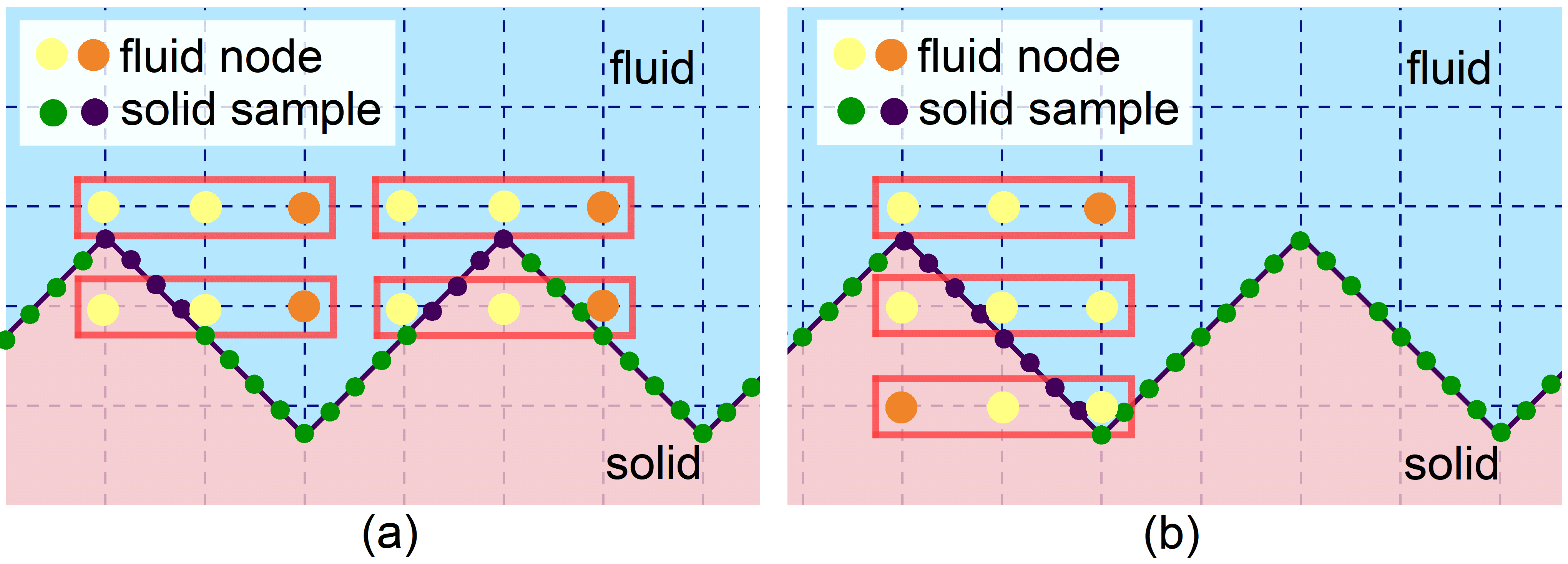}
	\vspace{-8mm}
	\caption{\textbf{2D illustration of caching for different block sizes.} Assume each warp has 8 threads.  In both (a) and (b), the black points represent solid samples processed by threads in one warp, with one thread assigned to each point.  When $\ell = 1$ as in (a), the points are dispersed, whereas for $\ell = 2$ as in (b) the points are more localized.  Now, assume 3 fluid nodes are stored in each memory segment. In (a) and (b), each rectangle marks the fluid nodes stored in one memory segment.  In (a), four memory segments need to be read to load the fluid data needed by the solid samples, while in (b) only three segments are needed due to greater sample locality.  This shows how varying $\ell$ can improve memory bandwidth and caching performance.}
		
	\label{fig:caching}
\end{figure}

To explain the utility of our proposed sample ordering for improving memory performance, we refer to Fig.~\ref{fig:caching} for a 2D illustration. Assume in this example that a GPU warp contains 8 threads.  In Fig.~\ref{fig:caching}(a) we use $\ell = 1$ whereas in Fig.~\ref{fig:caching}(b) we use $\ell = 2$.  The black points in both subfigures show the solid samples processed by one warp, with one thread assigned to each sample.  Note that due to the larger $\ell$, samples are more localized in Fig. 5(b).  Now, assume that 3 fluid nodes are stored in each segment of GPU memory.  Fig.~\ref{fig:caching}(a) shows that 4 memory segments, marked by the orange rectangles, need to be read to load the fluid data needed by the warp of solid samples when $\ell = 1$, whereas in Fig.~\ref{fig:caching}(b) only 3 memory segments need to be read for $\ell = 2$.  This shows that differences in the block size can cause variations, sometimes quite significant, in the amount of memory bandwidth consumed and cache hits obtained.

\subsection{Reducing Load Imbalance and Warp Divergence}
\label{sec:warp_divergence}
As stated earlier, the use of the IB method removes the conditional branching which occurs during boundary treatment in the bounce-back schemes used in many kinetic solvers.  However, a baseline implementation of the IB method requires different numbers of loop iterations for different fluid nodes around the solid boundary, with the variation often being quite large, and this leads to substantial load imbalance between the threads.  In this section, we propose additional changes to the traditional IB implementation to improve load balancing.  


\subsubsection{Load imbalance in the immersed boundary method}
We first show how a baseline implementation of the IB method can result in load imbalance among threads.  
Consider the force spreading step described in Sec. \ref{sec:boundarytreat} (cf. Eq.~\eqref{eq:IB-LBM-spreading}).  Here, threads are assigned to fluid nodes and the penalty force at a fluid node is computed by summing together all the penalty forces, computed immediately after interpolation by Eq.~\ref{eq:IB-LBM-interpolation}, at nearby solid samples (this step is also called ``gathering''~\cite{heron1989particle}), weighted by a kernel $K$ in a 2$\times$2$\times$2 neighborhood in 3D (2$\times$2 in 2D), as shown in the blue box in Fig.~\ref{fig:ib-method} for the 2D case.
While Poisson-disk sampling ensures almost uniform samples over the solid surface, each 2$\times$2$\times$2 neighborhood may contain local surfaces with different areas due to variations in local geometry, resulting in large variance in the number of samples in the force summation at different fluid nodes (see Table.~\ref{tab:gathering_scattering} for the variation in the number of loop iterations in an example computation).  This problem is especially severe when the solid geometry is complex and local geometric roughness changes with location.  


\subsubsection{Load balancing the immersed boundary method}
\label{sec:ib_warp_divergence}
While we saw above that parallelizing across fluid nodes can result in substantial load imbalance, the imbalance can be significantly reduced if we instead parallelize over the solid samples.  This technique is also called ``scattering'', and has been used in other GPU-based particle simulations, e.g.~\cite{heron1989particle}.  In particular, we assign one GPU thread to process each solid sample.  In the interpolation process, each sample interpolates fluid velocities from nearby grid nodes, while in the force spreading process, each solid sample redistributes its penalty force to nearby fluid nodes, as shown in the red box in Fig.~\ref{fig:ib-method}.
Eq.~\eqref{eq:IB-LBM-spreading} is equivalent to adding a penalty force with the corresponding kernel weight $K(\|\bm{x}_s - \bm{x}_f^i\|) \bm{g}^{s \rightarrow f}(\bm{x}_s)$ to the fluid velocities of all grid nodes $\bm{u}(\bm{x}_f^i)$ in the same 2$\times$2$\times$2 (or 2$\times$2) neighborhood.
While the number of fluid nodes still varies for each solid sample, the variation is much less than the variation in the number of solid samples in each neighborhood, leading to greatly reduced load imbalance (see Table.~\ref{tab:gathering_scattering} for the reduction in load imbalance using scattering).  Note that since multiple solid samples may try to add forces to the same fluid node, we use atomic additions to avoid race conditions.

\begin{table}
	\renewcommand{\arraystretch}{1.5}
	\setlength{\tabcolsep}{2.3mm}{
	\begin{tabular}{|c|c|c|c|}
		\hline 
		
		\multirow{2}{*}{} &
		\multicolumn{2}{c|}{number of loops} &
		\multirow{2}{*}{time (ms)} \\
		
		\cline{2-3}   
		& min. & max. &  \\
		\hline
		
		gathering with grid acceleration & 2 & 754 & 53.9 \\
		\hline
		scattering with atomic addition& 8 & 27 & 6.1 \\
		\hline
		
	\end{tabular}}
	\vspace*{2mm}
	\caption{Comparison of the number of loop iterations using ``gathering'' and ``scattering'', and the improvement in performance.}
	\vspace{-5mm}
	\label{tab:gathering_scattering}
\end{table}

\subsubsection{Domain boundary treatment}
\label{sec:domain_bc_treatment}
The method described above significantly reduces load imbalance for the IB method at solid boundaries.  However, special treatment is still needed for the domain boundary.  Since the IB method is difficult to apply at domain boundaries, at these locations we resort to the traditional bounce-back scheme.
Recall that the main drawback of this method is that it causes warp divergence due to conditional branching.  To mitigate this effect, we perform bounce-back using 6 different kernels for the 3D case (4 different kernels for 2D case), one for each face of the domain boundary.  For each face, threads only need to check one condition, namely whether they border the face being processed, while the distribution functions which need bounce-back treatment are known in advance and do not need to be determined using conditionals.
Overall, performance is improved using multiple kernel launches instead of a single one. 


\subsection{Increasing Occupancy}
\label{sec:occupancy}

GPU occupancy refers to the number of active GPU threads, which should be maximized in order to make full use of the processing capabilities of the GPU.
One factor which limits occupancy is the amount of resources, such as registers or shared memory, needed by each thread.  The ACM-MRT-based kinetic solver we employ produces significantly higher quality results compared to other kinetic methods.  However, this comes at the cost of needing to perform very lengthy arithmetic computations in the collision process, which in turn requires a large number of GPU registers and results in low occupancy.  Thus, to improve performance, it is important that we relieve register pressure.


\subsubsection{Using multiple kernel launches}
One straightforward strategy for reducing the number of required registers is to separate the most complex part of the collision computation, namely the calculation of the $\Omega_i$ ($i \in \{0, \ldots, 26\}$) values, into multiple kernel launches.
That is, each kernel launch only computes a portion of the $\Omega_i$ values.
As we increase the number of kernel launches, the registers needed for each thread decreases.
However, each kernel launch requires reading and writing partially computed $f_i$ values from and to the GPU main memory.
Thus, a tradeoff exists between the higher occupancy enabled by a larger number of kernel launches and the cost of a larger number of memory accesses.
In practice, we found that using two kernel launches typically achieved the best balance and highest performance in our GPU system.

\subsubsection{Simplification of relaxation computation}
In addition to separating the computation of different $\Omega_i$'s into multiple kernels, another strategy to reduce the number of required registers is to simplify the lengthy arithmetic expressions used in computing collision.
The specific form of the expression for one of the collision terms used in the ACM-MRT model is given in the \textit{supplementary document} and will not be covered in detail here.
However, a key step involves the computation of  $\bm{M}^{-1}$, an example of which is given by the expression shown below.  Note that the following formula only shows about one sixth of the complete expression for $\Omega_{26}$.
\begin{equation}
\begin{aligned}
\Omega_{26} &= \tilde{f}_{20}/8 - \tilde{f}_{16}/8 + \tilde{f}_{21}/8 + \tilde{f}_{22}/8 + \cdots \\
&+ (\tilde{f}_{24}\bm{u}_y)/4 - (\tilde{f}_{4}\bm{u}_z)/8 + (\tilde{f}_{10}\bm{u}_z)/16 \cdots\\
&+ (\tilde{f}_{25}\bm{u}_z)/4 + (\tilde{f}_{6}{\bm{u}_x}^2)/8 - (\tilde{f}_{11}{\bm{u}_x}^2)/16 \cdots\\
&- (\tilde{f}_{19}{\bm{u}_y}^2)/16 + (\tilde{f}_{4}{\bm{u}_z}^2)/8 - (\tilde{f}_{10}{\bm{u}_z}^2)/16 \cdots\\ 
&- (\tilde{f}_{7}{\bm{u}_x}^2{\bm{u}_z}^2)/12 + (\tilde{f}_{8}{\bm{u}_x}^2{\bm{u}_z}^2)/24 \cdots\\
&- (\tilde{f}_{16}\bm{u}_x\bm{u}_y)/2 + (\tilde{f}_{22}\bm{u}_x\bm{u}_y)/2 + (\tilde{f}_{4}\bm{u}_x\bm{u}_z)/4 \cdots\\
&+ (\tilde{f}_{21}\bm{u}_x\bm{u}_z)/2 + (\tilde{f}_{4}\bm{u}_y\bm{u}_z)/4 + (\tilde{f}_{5}\bm{u}_y\bm{u}_z)/4 \cdots\\
&- (\tilde{f}_{5}\bm{u}_x{\bm{u}_y}^2)/4 - (\tilde{f}_{6}{\bm{u}_x}^2\bm{u}_y)/4 - (\tilde{f}_{7}\bm{u}_x{\bm{u}_y}^2)/24 \\
&- (\tilde{f}_{7}{\bm{u}_x}^2\bm{u}_y)/24 + (\tilde{f}_{8}\bm{u}_x{\bm{u}_y}^2)/12 + (\tilde{f}_{8}{\bm{u}_x}^2\bm{u}_y)/12 \cdots\\
&- (\tilde{f}_{4}\bm{u}_x{\bm{u}_z}^2)/4 - (\tilde{f}_{6}{\bm{u}_x}^2\bm{u}_z)/4 + (\tilde{f}_{7}\bm{u}_x{\bm{u}_z}^2)/12\\
&+ (\tilde{f}_{7}{\bm{u}_x}^2\bm{u}_z)/12 - (\tilde{f}_{8}\bm{u}_x{\bm{u}_z}^2)/24 - (\tilde{f}_{8}{\bm{u}_x}^2\bm{u}_z)/24 \cdots\\
&- (\tilde{f}_{4}\bm{u}_y{\bm{u}_z}^2)/4 - (\tilde{f}_{5}{\bm{u}_y}^2\bm{u}_z)/4 - (\tilde{f}_{7}\bm{u}_y{\bm{u}_z}^2)/24 + \cdots\\
&- (\tilde{f}_{7}{\bm{u}_y}^2\bm{u}_z)/24 - (\tilde{f}_{8}\bm{u}_y{\bm{u}_z}^2)/24 - (\tilde{f}_{8}{\bm{u}_y}^2\bm{u}_z)/24 \cdots\\
\end{aligned}
\label{eq:Omega_26}
\end{equation}
Here, $\tilde{f}_i=[\bm{D}\bm{M}(\bm{f}-\bm{f}^{eq})]_i$, and $[\cdot]_i$ picks the $i$'th element out of a vector.  Computing this and similar expressions requires dozens of registers for each GPU thread.  This can be reduced with the observation that the expressions for different $\Omega_i$'s share many similar terms, with the only difference being the coefficients.  
In general, all the terms appearing in an expression can be written as the sum of monomials of the form $\omega(\rho)\prod_{\alpha=0}^{2}\bm{u}_\alpha^{p_\alpha}$ ($p_\alpha = 0,1,2$ is the order of the $\alpha$'th component).
Different $\Omega_i$'s may have different combinations of orders with different weights $\omega(\rho)$.
Within each $\Omega_i$, any term having the same order of $p_0, p_1, p_2$ but different weights $\omega(\rho)$ can be merged, as shown in Table~\ref{tab:collision_model_illustration}.
Across different $\Omega_i$'s, we first compute the values of the shared terms and store them.  Later, we simply read back the values needed for each $\Omega_i$.  

\begin{figure}[t]
	\centering
	\includegraphics[width=0.95\columnwidth]{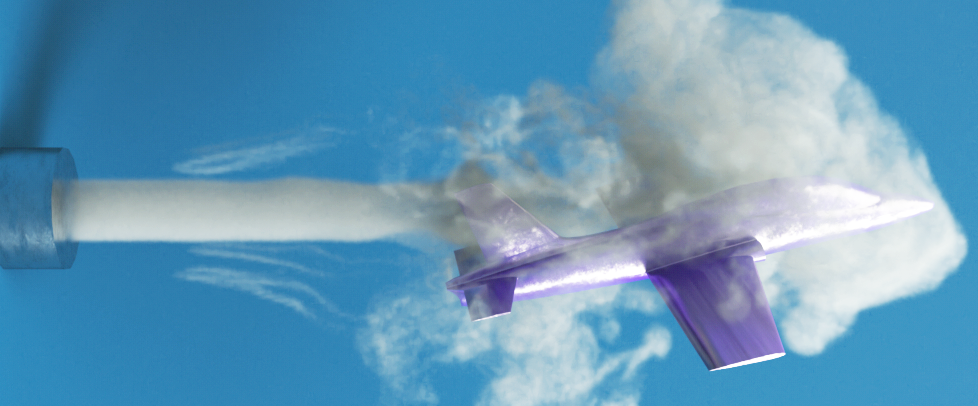}
	\vspace{-4mm}
	\caption{An example of a jet flow passing over a slanted airplane obstacle was used in the performance comparisons in Figs.~\ref{fig:streaming_performance} to \ref{fig:domain_boundary_performance}.}
	\label{fig:comparison-scene}
\end{figure}

\subsubsection{Discussion}
Based on our experiments for the test case shown in Fig.~\ref{fig:comparison-scene}, using a grid resolution of $100\!\times\!200\!\times\!100$ with 45,127 solid samples, collision performance is improved by around 35\% using a combination of multiple kernel launches and simplified collision terms, while arithmetic expression simplification alone improves performance by 6\%.

\begin{figure*}[t]
	\centering
	\includegraphics[width=0.9\textwidth]{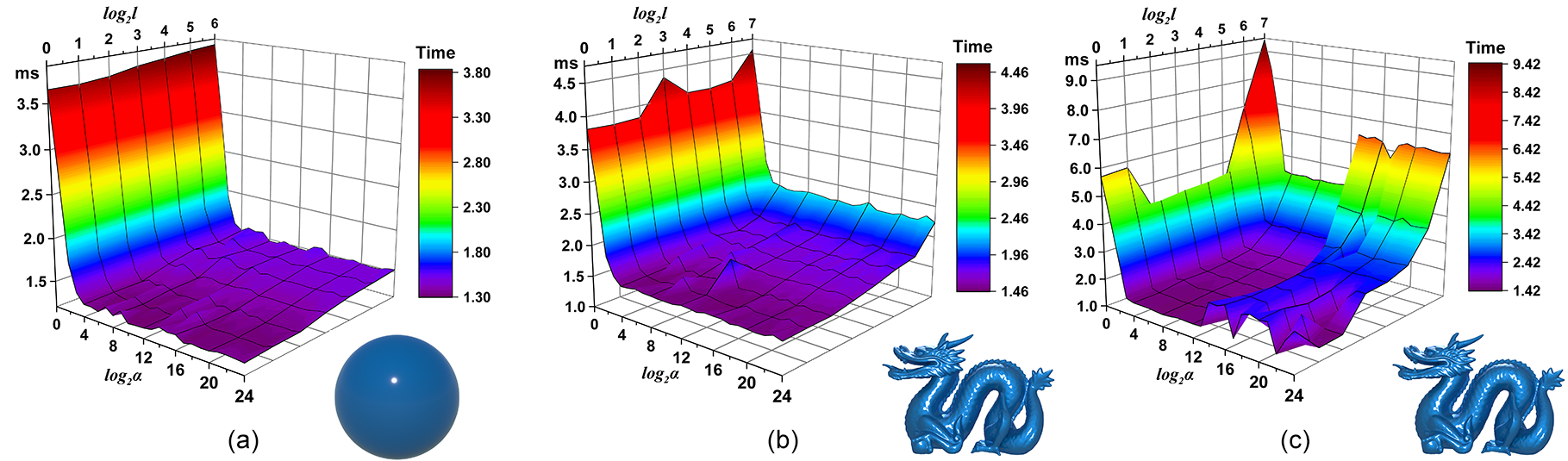}
	\vspace{-4mm}
	\caption{\textbf{Performance variations from changing model parameters for different solid geometries}. Given a simple geometry such as the ball in (a), the simulation cost does not change much as we vary the sample block size $\ell$ and the CSoA parameter $\alpha$ within a specific range ($\ell\in[2,4]$, $\alpha\geq64$).  However, for more complex geometries such as the dragon, the cost can vary significantly based on (b) the degree of concavity in the geometry compared to (a) and (c) the number of solid samples compared with (b) ((c) contains around 10 times more solid samples than (b)).  This indicates that searching for the optimal $\ell$ and $\alpha$ is important for complex geometries containing a large number of solid samples.}
	\label{fig:parameter-performance}
\end{figure*}

\begin{table}[t]
	\renewcommand{\arraystretch}{1.5}
	\setlength{\tabcolsep}{0.8mm}{
		\begin{tabular}{|c|c|c|c|c|c|}%
			\hline 
			& $\bm{u}_y$ & ... &
			${\bm{u}_x}{\bm{u}_y}^2$ &...& ${\bm{u}_x}^2{\bm{u}_y}^2{\bm{u}_z}^2$ \\[5pt]
			\hline
			$\Omega_0$ & $2(\tilde{f}_{12}-\tilde{f}_{25})$ & ... & $-\tilde{f}_{10}+\tilde{f}_{13}$ & ...& $-\rho$ \\[3pt]
			\hline 
			$\Omega_1$ & $-\tilde{f}_{5}- ... +\tilde{f}_{25}$  & ... & $-0.5\rho+...-0.5\tilde{f}_{13}$ &...& $0.5\rho$ \\[3pt]
			\hline 
			... & ... & ... & ... & ... & ...\\
			\hline 
			$\Omega_{13}$ & $0.5\tilde{f}_{5}- ... -0.5\tilde{f}_{25}$ & ... & $0.5\tilde{f}_{5}+...+0.25\tilde{f}_{13}$  & ... & $-0.25\rho$\\[3pt]
			\hline 
			$\Omega_{14}$ & $-0.5\tilde{f}_{5}+ ... -0.5\tilde{f}_{25}$ & ... & $0.5\tilde{f}_{5}+...+0.25\tilde{f}_{13}$ & ... & $-0.25\rho$\\[3pt]
			\hline 
			... & ... & ... & ... & ... & ...\\
			\hline
	\end{tabular}}
	\vspace*{2mm}
	\caption{Example of monomials and their corresponding weights used in computing $\bm{M}^{-1}$ in the ACM-MRT collision model.}
	\vspace{-7mm}
	\label{tab:collision_model_illustration}
\end{table}

\subsection{Our Parametric Model}
\label{sec:parametric_model}
In the preceding subsections, we described a number of different GPU optimizations for improving the performance of our kinetic solver.  While the optimizations may appear to be largely independent, they are in fact interconnected.
In particular, the solver's performance is strongly affected by the memory layout, which in turn is determined by the parameters $\ell$, giving the size of blocks that solid samples are grouped into, and $\alpha$, giving the granularity of the CSoA layout for fluids.
We therefore combine these parameters into a metric $H^t(\ell,\alpha)$ measuring the cost to simulate a time step starting from time $t$.  Note that $H^t(\ell,\alpha)$ may vary for different $t$, as the immersed solid may lie at different orientations over time, leading to different costs to implement the IB method.  To obtain more stable measurements, we consider measuring the average cost over a time interval $[t_s, t_e]$.
We frame this as searching for optimal values for $\ell$ and $\alpha$ such that the following cost is minimized:
\begin{equation}
\text{argmin}_{\ell, \alpha}\;\frac{1}{N_{t_s \rightarrow t_e}}\sum_{t=t_s}^{t_e} H^t(\ell,\alpha),
\label{eq:paramtric_model_single}
\end{equation}
where $N_{t_s \rightarrow t_e}$ is the number of time steps between $t_s$ and $t_e$.  Note that we may use different numbers of time steps in order to obtain optimal parameters balancing different cases in a simulation.  
For static solids, $N_{t_s \rightarrow t_e}$ is typically small (e.g. $N_{t_s \rightarrow t_e}\!=\!10$), while for dynamic solids, $N_{t_s \rightarrow t_e}$ should be large enough to cover many different orientations of the solid.  
Since translation of solid usually does not strongly affect the objective, we consider $N_{t_s \rightarrow t_e}$ only when rotating the solid object by an angle of $2\pi$ in 3D, leading to a value of around 500 for $N_{t_s \rightarrow t_e}$. 
This minimization is performed once before the start of every simulation with a different setting for the grid resolution, geometry shape, etc.


To minimize Eq.~\eqref{eq:paramtric_model_single}, we enumerate all possible combinations of $\ell$ and $\alpha$ and search for the combinations with minimal cost.
Since $\ell$ is the side length of the subdivided block, it is upper bounded by the smallest edge length $L_m$ of the solid's bounding box, and thus we search for $\ell \in \{1, 2, \ldots, L_m\}$.  For the CSoA parameter $\alpha$, we only consider powers of 2 since we want $\alpha$ to be divisible by or divisible into the thread warp size of 32.  We thus search for $\alpha \in \{2^1, 2^2, \ldots, 2^{\lfloor \log_2 N_f \rfloor}\}$, where $N_f$ is the total number of fluid nodes.  

Fig.~\ref{fig:parameter-performance} shows the simulation cost in seconds for certain values of $\ell$ and $\alpha$ and for different types of geometries.  We note that after minimizing Eq.~\eqref{eq:paramtric_model_single}, we can obtain the best parameters for the scenario we want to simulate.  But if the scenario changes, e.g. if we use a different size for the fluid domain, a different solid geometry, or if the solid is sampled with a different number of points, we should perform the minimization again.  The cost of the procedure is usually not excessive, e.g. 5 to 15 minutes at a grid resolution of $200\!\times\!400\!\times\!200$, compared to a total simulation time of several hours.
As the simulator can sometimes be several times faster using a good parameter setting compared to a poor one, it is usually well worthwhile to perform optimal parameter search before starting a simulation.


\begin{algorithm}[h]
	\small
	\caption{Our GPU-optimized IB-based kinetic solver}\label{alg:opti_gpu_lbm} 
	\begin{algorithmic}
		\State  \textit{// Find optimal parameters for $\ell$ and $\alpha$.}
		\State SearchOptimalParameters($\ell$, $\alpha$); \Comment{Sec.~\ref{sec:parametric_model}}
		\State ReorganizeMemory($\ell$, $\alpha$); \Comment{Eq. \eqref{eq:indexing}} for arrays of vectors\\
		
		\While {time step $\leq$ maximum time step}
		\State \textit{//} (a): \textit{Perform parallel streaming.} \Comment{Sec.~\ref{sec:straight_implementation}}
		\State \textbf{parallel for} each fluid node $\bm{x}_f$ \textbf{do}
		\State \qquad \textbf{for} each $i$ $\in [0,26]$ of node $\bm{x}_f$ \textbf{do}
		\State \qquad \qquad $j$ = RemapIndex($\bm{x}_f$, $i$); \Comment{Eq.~\eqref{eq:indexing}}
		\State \qquad \qquad $j_n$ = RemapIndex($\bm{x}_f-\bm{c}_i$, $i$); \Comment{Eq.~\eqref{eq:indexing}}
		\State \qquad \qquad $f^{t}$[$j$] = $f^{t-1}$[$j_n$];\Comment{Eq.~\eqref{eq:lbm_streaming}}
		\State \qquad \textbf{end for}
		\State \textbf{end for}\\
		
		\State \textit{//} (b): \textit{Compute macroscopic variables in parallel.} \Comment{Sec.~\ref{sec:straight_implementation}}
		\State \textbf{parallel for} each fluid node $\bm{x}_f$ \textbf{do}
		\State \qquad UpdateDensityAndVelocity($\bm{x}_f$);  \Comment{Eq.~\eqref{eq:discrete_lbm_moments}}
		\State \qquad ResetForce: $\bm{g}[\bm{x}_f]$ = $\bm{0}$;
		\State \textbf{end for}\\
		
		\State \textit{//} (c): \textit{Immersed boundary treatment.} \Comment{Sec.~\ref{sec:warp_divergence}}
		\State \textbf{parallel for} each solid sample $\bm{x}_s$ \textbf{do}
		\State \qquad ComputePenaltyForce($\bm{x}_s$); \Comment{Eq.~\eqref{eq:IB-LBM solid-to-fluid}}
		\State \textbf{end for}
		\State \textbf{parallel for} each solid sample $\bm{x}_s$ \textbf{do}
		\State \qquad SpreadPenaltyForce($\bm{x}_s$);  \Comment{Eq.~\eqref{eq:IB-LBM-spreading} (atomic addition)} 
		\State \textbf{end for} 
		\State ParallelTotalForceAndTorqueCalculation();
		\State ParallelSolidSampleUpdates();\\
		
		\State \textit{//} (d): \textit{Perform parallel collision.} \Comment{Sec.~\ref{sec:occupancy}}
		\State \textbf{parallel for} each fluid node $\bm{x}_f$ \textbf{do}
		\State \qquad Compute $\bm{m}$[$\bm{x}_f$] and $\bm{m}^{eq}$[$\bm{x}_f$];
		\State \textbf{end for}
		\State \textbf{parallel for} each fluid node $\bm{x}_f$ \textbf{do}
		\State \qquad Compute $\Omega _{0-13}$[$\bm{x}_f$] and add $G_{0-13}$[$\bm{x}_f$]; \Comment{Eq.~\eqref{eq:lbm-mrt-relaxation-moments}}
		\State \textbf{end for}
		\State \textbf{parallel for} each fluid node $\bm{x}_f$ \textbf{do}
		\State \qquad Compute $\Omega_{14-26}$[$\bm{x}_f$] and add $G_{14-26}$[$\bm{x}_f$]; \Comment{Eq.~\eqref{eq:lbm-mrt-relaxation-moments}}
		\State \textbf{end for}\\
		
		\State \textit{//} (e): \textit{Process domain boundary in parallel.} \Comment{Sec.~\ref{sec:domain_bc_treatment}}
		\State ParallelLeftBoundaryTreatment();
		\State ParallelRightBoundaryTreatment();
		\State ParallelUpBoundaryTreatment();
		\State ParallelDownBoundaryTreatment();
		\State ParallelForwardBoundaryTreatment();
		\State ParallelBackBoundaryTreatment();
		\EndWhile
	\end{algorithmic}
\end{algorithm}

\begin{figure}[t]
	\centering
	\includegraphics[width=0.95\columnwidth]{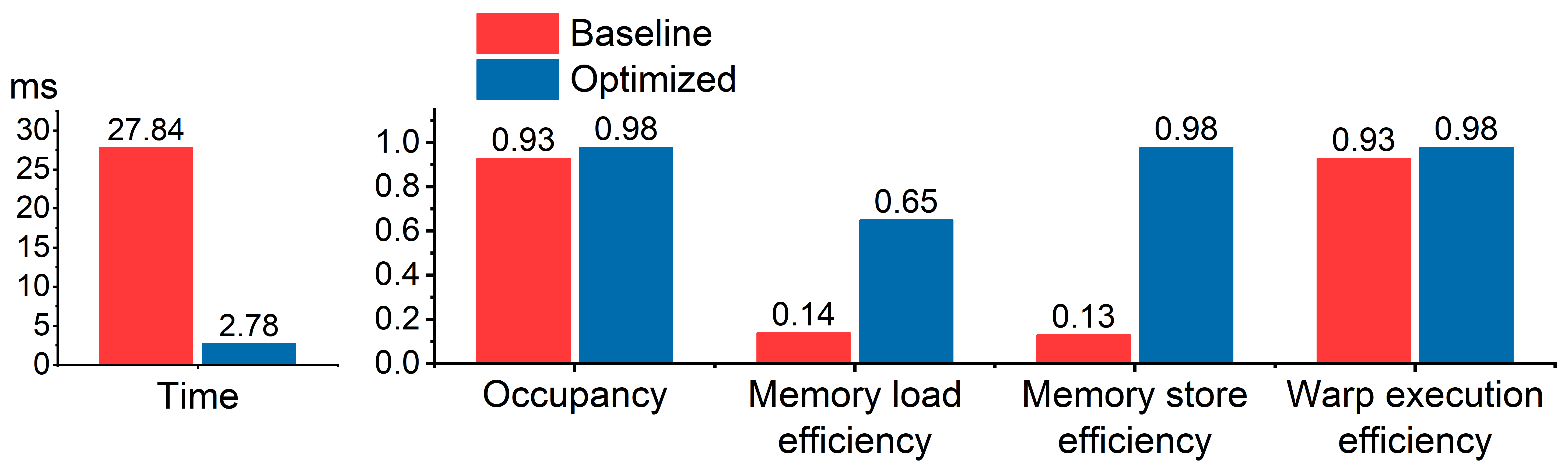}
	\vspace{-4mm}
	\caption{Comparison of streaming performance with (blue bar) and without (red bar) our optimizations.}
	\label{fig:streaming_performance}
\end{figure}

\begin{figure}[t]
	\centering
	\includegraphics[width=0.95\columnwidth]{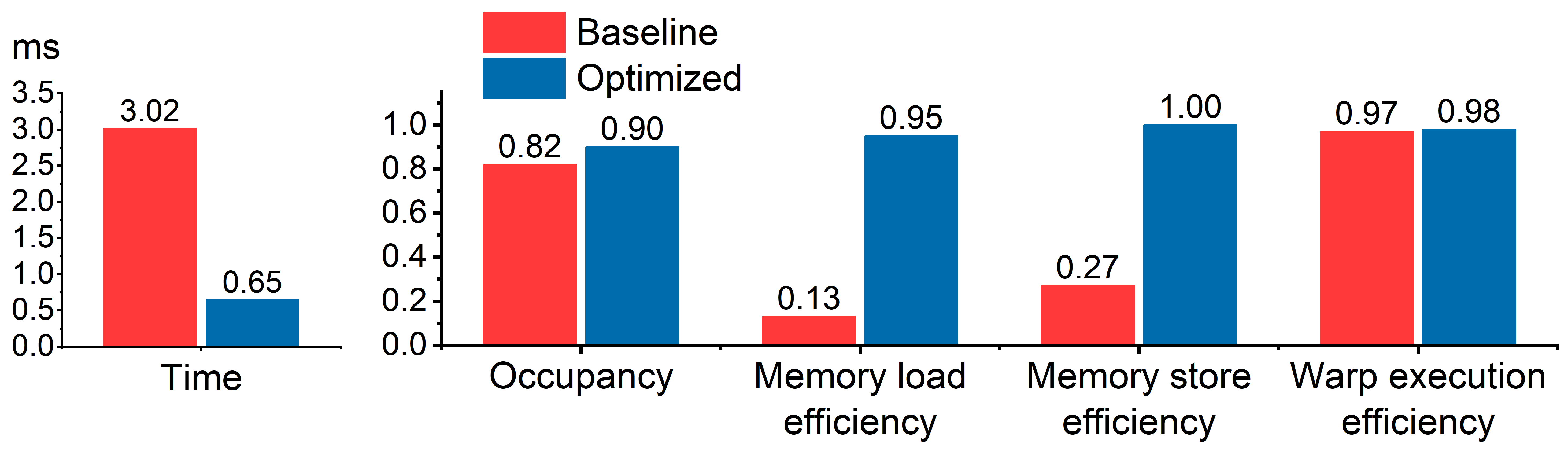}
	\vspace{-4mm}
	\caption{Comparison of performance for calculating macroscopic variables with (blue bar) and without (red bar) our optimizations.}
	\label{fig:macroscopic_variable_performance}
\end{figure}

\subsection{Performance Analysis}
\label{sec:performance_analysis}
We summarize our entire optimized kinetic solver in Alg.~\ref{alg:opti_gpu_lbm}.  In the rest of this section, we analyze the performance of different parts of the solver.
We compared the performance of our optimized solver with the baseline solver described in Sec.~\ref{sec:straight_implementation}, which is for the most part a faithful implementation of the algorithm described in \cite{Li-2018}, except that \cite{Li-2018} did not use the immersed boundary method.  To make our analysis as detailed as possible, we decomposed our solver into its main subroutines (cf. Sec. \ref{sec:straight_implementation}), consisting of streaming, macroscopic fields calculation, the IB method, collision, and domain boundary treatment.  We analyzed each of these components using key GPU performance metrics including thread occupancy, memory load/store efficiency, warp divergence and warp execution efficiency.

\begin{figure}[t]
	\centering
	\includegraphics[width=0.95\columnwidth]{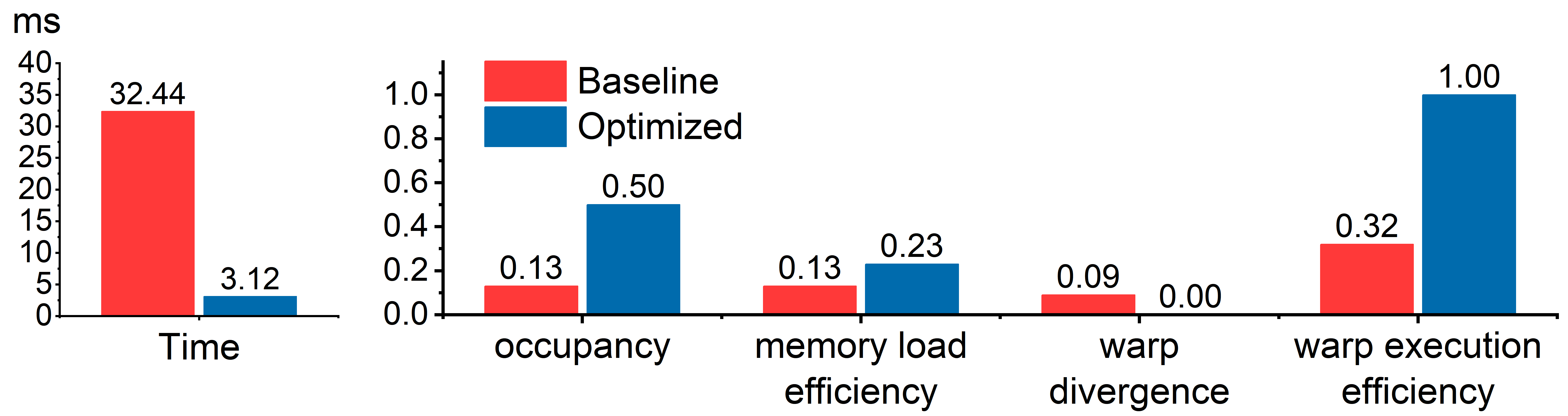}
	\vspace{-4mm}
	\caption{Comparison of immersed boundary computation performance with (blue bar) and without (red bar) our optimizations.}  
	\label{fig:ib-method_performance}
\end{figure}

\begin{figure}[t]
	\centering
	\includegraphics[width=0.95\columnwidth]{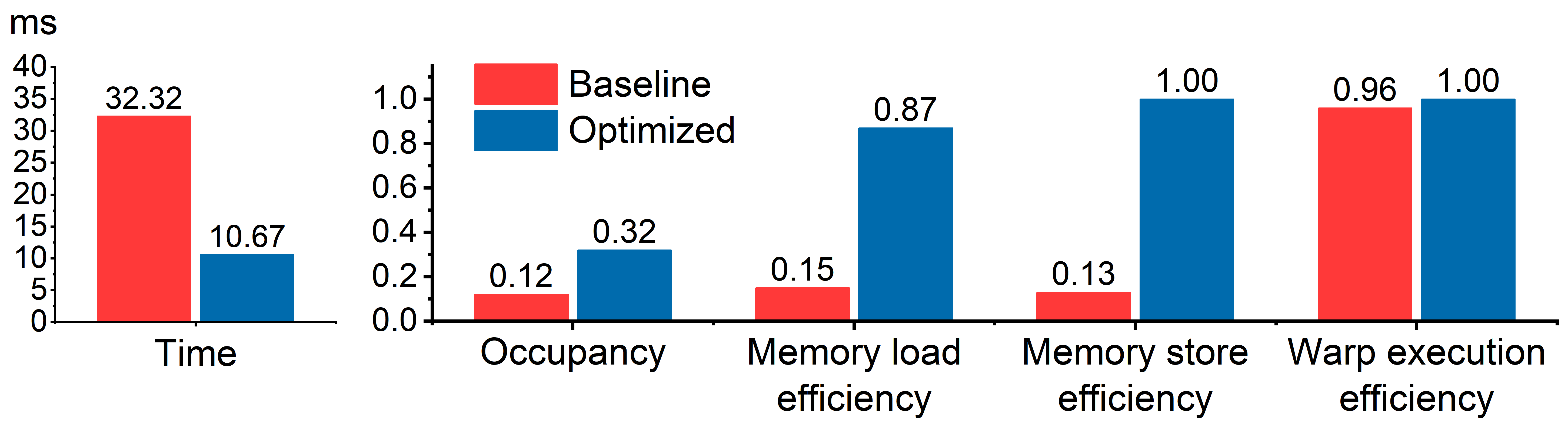}
	\vspace{-4mm}
	\caption{Comparison of collision performance with (blue bar) and without (red bar) our optimizations.}
	\label{fig:collision_performance}
\end{figure}

\begin{figure}[t]
	\centering
	\includegraphics[width=0.95\columnwidth]{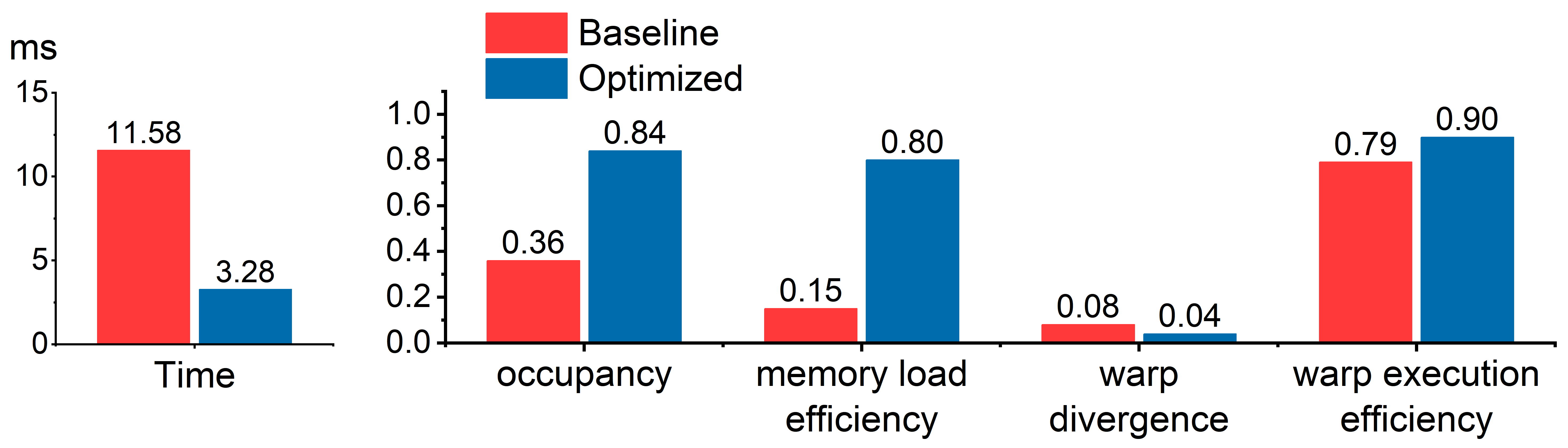}
	\vspace{-4mm}
	\caption{Comparison of domain boundary performance with (blue bar) and without (red bar) our optimizations.}
	\label{fig:domain_boundary_performance}
\end{figure}

Figs.~\ref{fig:streaming_performance} to \ref{fig:domain_boundary_performance} show performance comparisons of different parts of our optimized solver and the baseline solver using the scene setup shown in Fig.~\ref{fig:comparison-scene}.  We see that our parameterized data layout significantly improves memory access efficiency.  Occupancy was also improved several fold during the IB and collision computations, while warp execution efficiency was improved by a factor of three during the IB computation.  The overall performance was improved by a factor of 5 to 10 compared to the straightforward baseline implementation.   

\subsection{Extension to multi-GPU systems}
The optimizations described in the previous sections were performed on a single GPU, limiting the resolution at which the kinetic solver can simulate.
To support high resolution scenarios for large-scale simulations, we modify the algorithm to run in parallel on multiple GPUs.
Since the kinetic solver uses only local information from one-hop neighbors, we can also easily extend the optimizations techniques described earlier to multi-GPU systems.

\begin{figure}[t]
	\centering
	\includegraphics[width=\columnwidth]{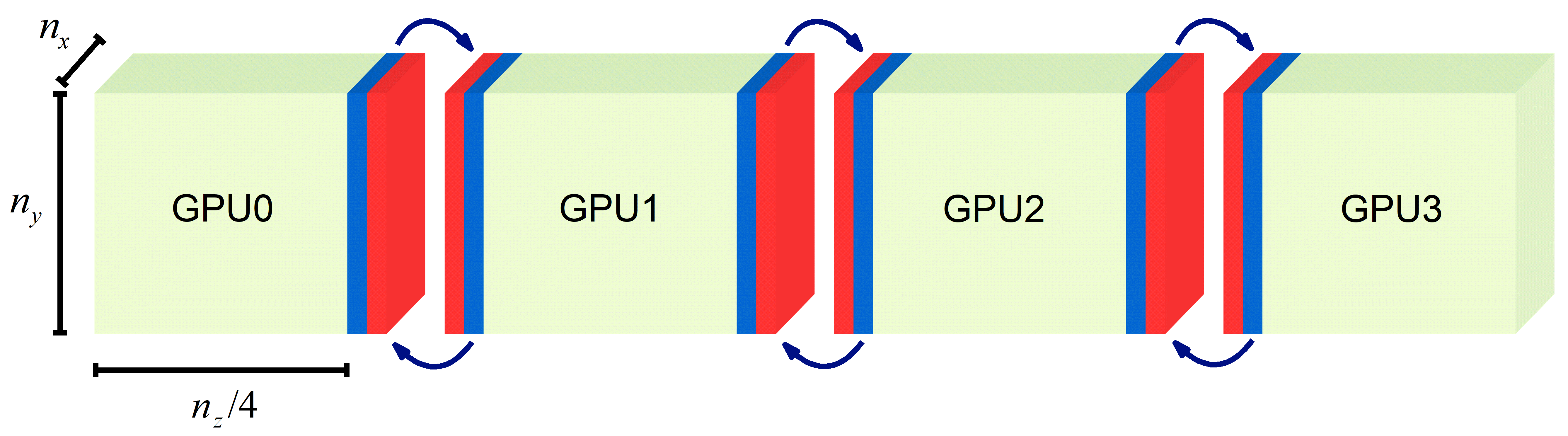}
	\vspace{-4mm}
	\caption{\textbf{Multi-GPU simulation.} The whole fluid domain is subdivided into $m$ ($m\!\leq\!4$ in our case) regions of roughly equal size along the $z$-dimension. We extend the shared face between neighboring GPUs by one cell along the $z$-direction and copy data from the neighboring GPUs (the blue regions), with arrows indicating the direction of copying. $n_x$, $n_y$ and $n_z$ are numbers of fluid grid nodes in the $x$-, $y$- and $z$-directions.}
	\label{fig:multi_gpu}
\end{figure}

We consider a setting where $m$ GPUs (in our case $m\!=\!4$) are installed on a single cluster node (the multi-node case is more complex and we leave it for future work).  We first subdivide the fluid nodes in the simulation domain into multiple regions along one dimension (e.g. the $z$-dimension), as shown in Fig.~\ref{fig:multi_gpu}, while ensuring that each region contains roughly the same number of fluid nodes.
For each pair of neighboring regions, we extend both regions by a one-cell-wide ``ghost region'' (marked in red in Fig. \ref{fig:multi_gpu}), holding data from the shared face (marked in blue) with its neighbors.  Data is updated (copied) between neighboring GPUs after every simulation time step.  This communication can be done using GPU to GPU memory transfer when supported by hardware.  However, due to limitations of our current hardware and for the sake of implementation portability, our current implementation performs data transfer through the CPU.  Note that this may substantially reduce performance, especially for large $m$, and that with faster inter-GPU communication mechanisms all the results described below are expected to improve.  The optimizations described earlier can be performed independently within each region.  When immersed solid does not move, we can subdivide solid samples into $m$ groups based on regions they lie in.  However, when the solid is allowed to move, its sample points may reside in different regions over time.  To deal with this situation, we store all solid samples in every GPU, thus avoiding the need to copy solid samples among GPUs, while incurring the cost of a slight increase in memory usage.  

%% file: 6.results-and-discussion.tex
\section{Results and Discussions}
\label{sec:results_discussions}
We implemented the single GPU version of our solver on an NVIDIA TITAN XP GPU with 12 GB of memory, installed on an 8-core Intel Xeon E5-2650 workstation with 64 GB of memory.
For the multi-GPU version of our algorithm, we used a system with 4 NVIDIA Tesla P40 GPUs per node, each having 24 GB of memory.  The system also had 4 Intel Xeon E5-2620 CPUs (8 cores per CPU) with 384 GB of memory.  Optimizing model parameters, as described in Sec. \ref{sec:parametric_model}, usually took 2 to 5 minutes for flows with static solids and 5 to 15 minutes for flows with dynamic solids at a grid resolution of $200\times400\times200$.
To render the simulated flows, we usually injected smoke particles, which were traced in parallel on the GPU.
We then projected the particle densities into a uniform volume grid, and used Mitsuba~\cite{JAKOB-2010} to perform high quality rendering.
The specific parameter settings for the simulations shown in this paper and the related performance statistics are reported in Table~\ref{tab:parameter-time}.

\begin{table*}[t]
	\centering
	
	\begin{center}
	\scalebox{1}
	{
		\begin{tabular}{lccccccccc}
			Figures. & Grid resolution &No. of solid samples &$\ell$&$\text{log}_2\alpha$& No. of GPUs & Storage & time / $\Delta t$ & total time steps & total time cost\\
			Fig.~\ref{fig:teaser} &  1200$\times$250$\times$840  &3,688,157& 2 &28& 4 & 58.4 GB & 1.21 sec.&76,600&1544.8 min.\\
			Fig.~\ref{fig:smoke-collide} (a) & 400$\times$200$\times$400  &48,336& 2 &9& 1  & 7.4 GB &0.56 sec.&20,000&186.7 min.\\
			Fig.~\ref{fig:smoke-collide} (b) & 400$\times$200$\times$400 &53,104& 4&16& 1  & 7.4 GB &0.58 sec.&25,600&247.5 min.\\
			Fig.~\ref{fig:high-res-result-car} & 2240$\times$320$\times$480 &1,971,941&2&29& 4  & 79.5 GB &1.70 sec.&26,300&745.2 min.\\
			Fig.~\ref{fig:high-res-airplane} &2100$\times$240$\times$600 &909,788&2&28& 4  & 69.9 GB &1.43 sec.&41,500&989.1 min.\\
		\end{tabular}
	}

	\end{center}

	\vspace*{1mm}
	\caption{Parameter settings and performance statistics for different simulations shown in the paper.}
	\vspace{-7mm}
	\label{tab:parameter-time}
\end{table*}


\begin{figure}[t]
	\centering
	\includegraphics[width=\columnwidth]{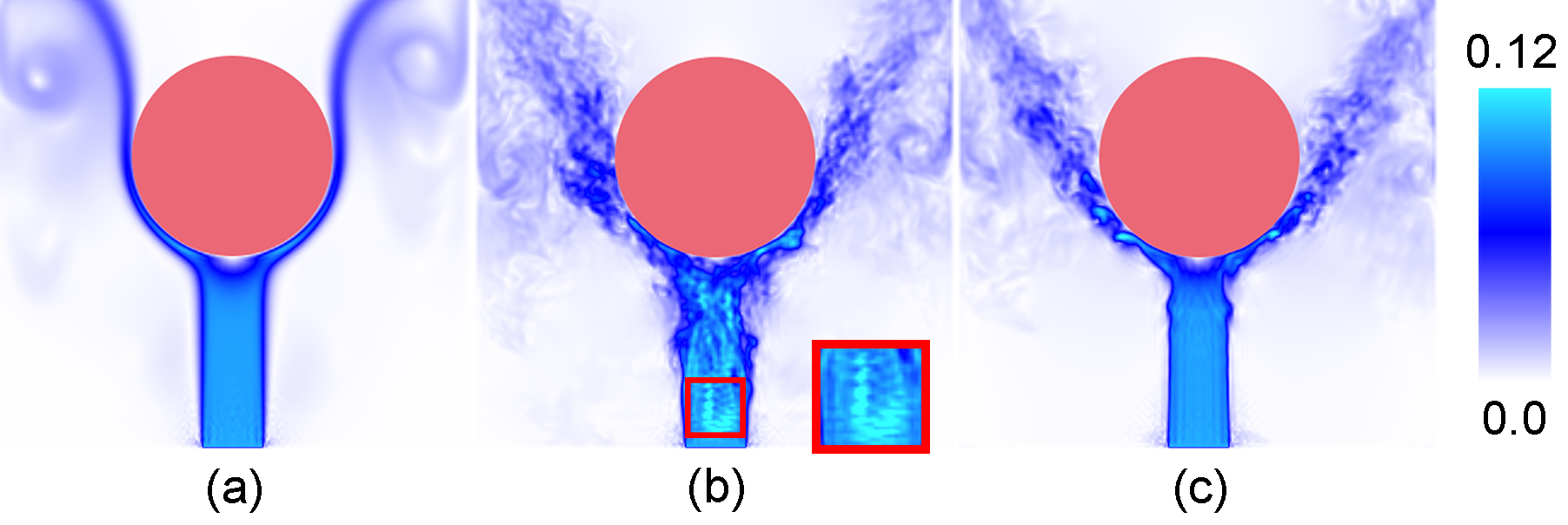}
	\vspace{-6mm}
	\caption{\textbf{Accuracy comparison for kinetic solvers using different collision models}. Visualization of velocity magnitude from a cross-section of a 3D velocity field for a jet flow impinging on a ball obstacle: (a) using the BGK model with the smallest stable viscosity ($2\!\times\!10^{-3}$), (b) using the MRT model with the smallest stable viscosity ($10^{-4}$), and (c) using the ACM-MRT model with the same viscosity as (b). The color bar shows the ranges of velocity magnitudes in LBE scale.}
	\label{fig:LBM_comparison}
\end{figure}

\subsection{Quality comparison of GPU-based kinetic solvers}
Recent GPU optimized kinetic solvers have typically employed the BGK or RM-MRT~\cite{suga2015d3q27} collision models.
In this section, we compare the quality of simulations obtained using these models with that using the ACM-MRT model.  We also compare the performance of these models.  

To ensure fairness in the quality comparisons, we use the D3Q27 lattice for all three collision models.  Fig.~\ref{fig:LBM_comparison} shows velocity fields obtained from the D3Q27 BGK, RM-MRT and ACM-MRT collision models.  Since the BGK model is not stable at high Reynolds numbers, we set its Reynolds number as high as possible while remaining stable.  Despite this, we still obtain an unnaturally smooth result, as shown in Fig. \ref{fig:LBM_comparison} (a).
The RM-MRT model can be stable at high Reynolds numbers but will be very dispersive, as seen in Fig.~\ref{fig:LBM_comparison} (b), and exhibit unnatural oscillation artifacts, as shown in the magnified inset.  Finally, the ACM-MRT model can support very high Reynolds numbers with very limited diffusion and dispersion, and produces the most visually appealing result, as shown in Fig.~\ref{fig:LBM_comparison} (c).

When comparing performance, we first note that to our knowledge, there are no GPU-optimized kinetic solvers using the D3Q27 lattice structure, and also no GPU-based solvers using the immersed boundary method.  For these reasons, it is difficult to find suitable existing works against which to compare the performance of our optimized solver.
Instead, we have chosen to implement D3Q27 versions of the BGK and RM-MRT models using an optimized CSoA data layout to compare with our solver.  We now give a detailed analysis showing the performance advantages of our solver in both single and multi-GPU settings for various resolutions of the simulation domain.  

\begin{figure}[t]
	\centering
	\includegraphics[width=0.95\columnwidth]{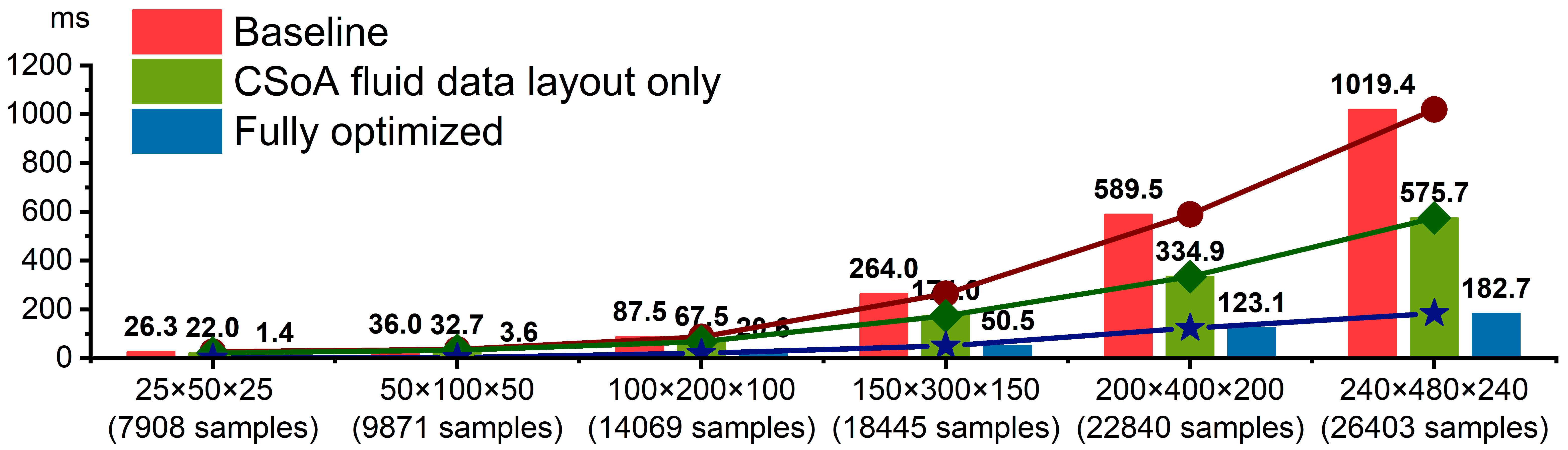}
	\vspace{-4mm}
	\caption{\textbf{Single GPU performance comparisons.} Red bars: the baseline kinetic solver (cf. Sec. \ref{sec:straight_implementation}).  Green bars: the modified baseline kinetic solver using CSoA data layout for fluid nodes.  Blue bars: our GPU-optimized kinetic solver. The comparisons were done with simulations at different grid resolutions and with resolution-matched numbers of solid samples in order to maintain simulation accuracy.}
	\label{fig:LBM_different_resolution}
\end{figure}

\subsubsection{Single GPU comparison}
We first compare the performance of our solver on a single GPU with the baseline solver described in Sec. \ref{sec:straight_implementation}, and a solver using an optimized CSoA data layout for fluid nodes, but employing none of the other optimizations described in Sec. \ref{sec:method}.
Fig. \ref{fig:LBM_different_resolution} shows the performance of the three solvers at different grid resolutions.  The baseline implementation suffers from uncoalesced memory accesses, and its cost increases significantly as resolution increases.
The CSoA-based solver is able to coalesce memory accesses for fluid nodes but does not optimize the IB method, so it also shows limited performance as the resolution increases.  Finally, our optimized solver shows increasing performance improvements relative to the other two solvers as resolution increases, and is over $3.1\times$ and $5.5\times$ faster than the CSoA and baseline solvers at the highest resolution we tested.

\begin{figure}[t]
	\centering
	\includegraphics[width=0.95\columnwidth]{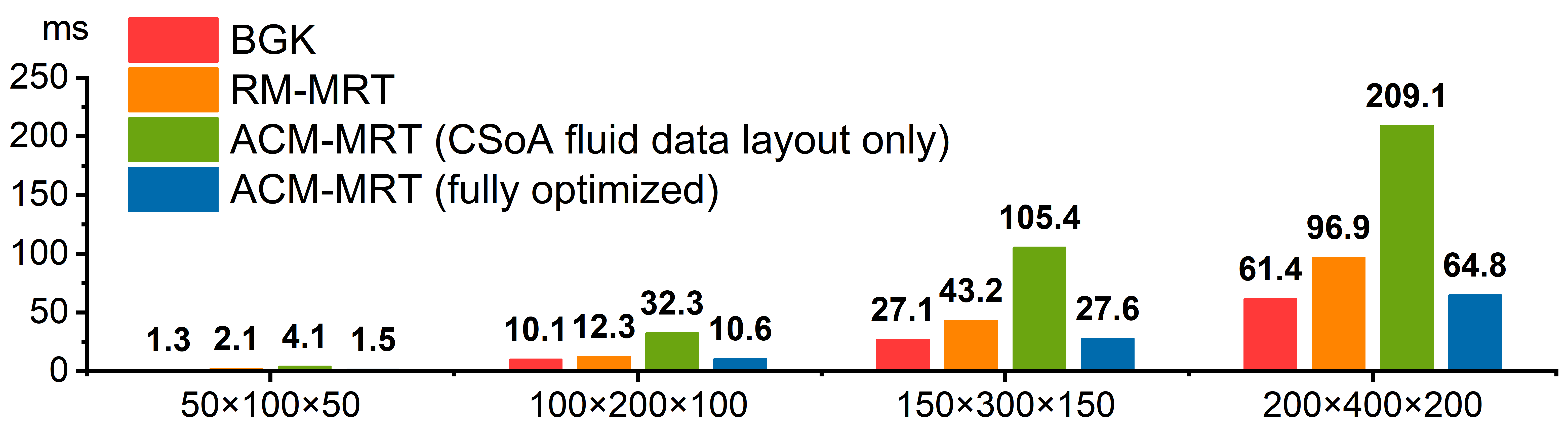}
	\vspace{-4mm}
	\caption{\textbf{Performance comparisons for different collision models.} We show timings for the collision step only, as the other parts of the different models employ the same optimizations and thus have nearly the same performance.  At different resolutions, our GPU-optimized ACM-MRT model performs nearly as efficiently as the basic BGK model, but produces much more accurate results for turbulent flows.}
	\label{fig:collision_comparison}
\end{figure}

We also present another comparison of the BGK, RM-MRT and ACM-MRT implementations in which we employed the same optimizations described in Sec. \ref{sec:method} for all three models, and only vary how the collision operation is performed.  The results are shown in Fig.~\ref{fig:collision_comparison}.  The plot shows that our optimized ACM-MRT implementation has nearly the same computational efficiency as the significantly simpler BGK model, while producing much more visually appealing outputs.  In general, the cost of arithmetic calculations occupies a relatively small portion of the overall running time, indicating that efficient memory accesses play a key role in optimizing kinetic solvers on the GPU.

\subsubsection{Multi-GPU comparison}
In this section, we analyze the performance of our optimized kinetic solver running on different numbers of GPUs at the same grid resolution.  As stated earlier, communication among GPUs is currently done by copying data through the CPU, and we expect that direct GPU to GPU communication can lead to substantial performance improvements.  

\begin{figure}[t]
	\centering
	\includegraphics[width=\columnwidth]{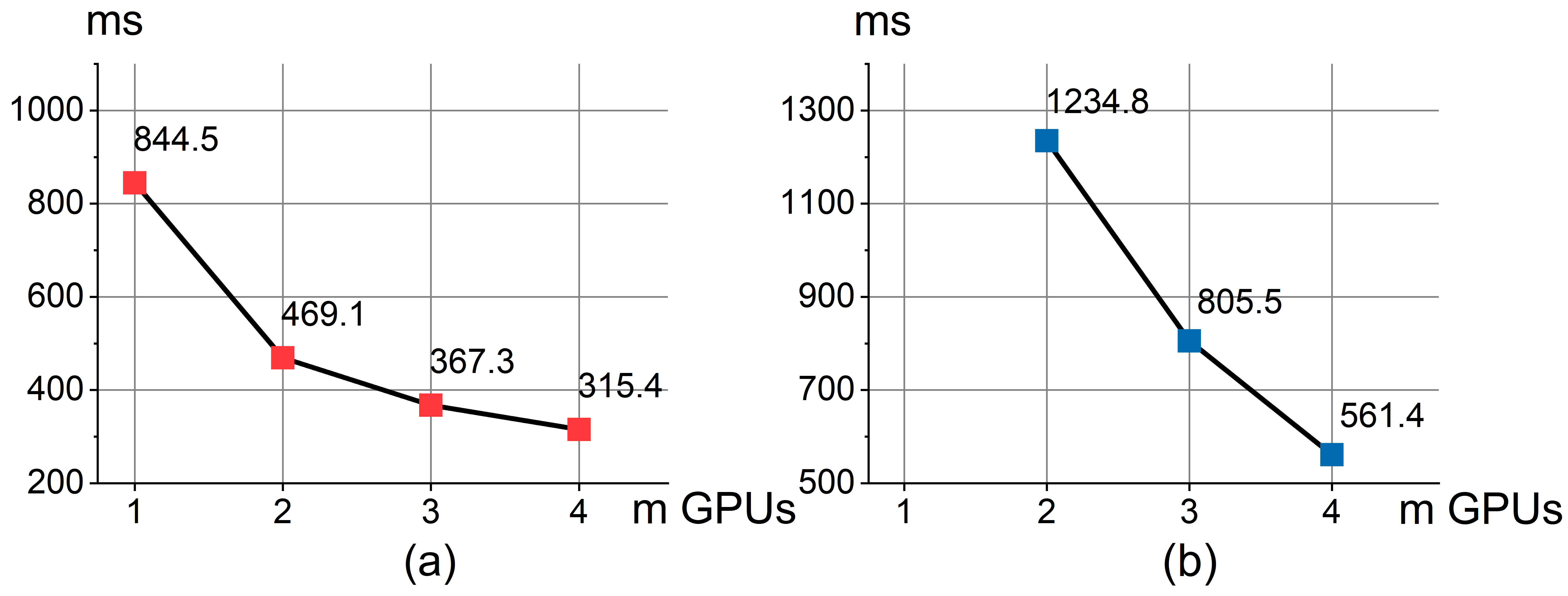}
	\vspace{-5mm}
	\caption{\textbf{Performance comparison for multi-GPU simulations}. In (a), we use a grid resolution of $200\!\times\!400\!\times\!200$ and 1,579,468 solid samples on 1 to 4 GPUs.  In (b), we increase the simulation size to $400\!\times\!400\!\times\!400$ with the same number of solid samples, and run on 2 to 4 GPUs.  Data copying has a larger impact on the smaller simulation, while the higher-resolution simulation shows improved scalability.}
	\vspace{-5mm}
	\label{fig:amount_of_gpu}
\end{figure}

Fig.~\ref{fig:amount_of_gpu} (a) shows the change in runtime when we perform a simulation of size $200\!\times\!400\!\times\!200$ using 1 to 4 GPUs.  We observed nearly perfect scaling when going from 1 to 2 GPUs.  However, as the number of GPUs increases, the communication cost and serialization at the CPU start to dominate, leading to reduced scaling.  To better understand the impact of communication on performance, we also performed a simulation of size $400\!\times\!400\!\times\!400$ on 2 to 4 GPUs as shown in  Fig.~\ref{fig:amount_of_gpu} (b).  Since we subdivided the simulation domain along the $z$-axis, the amount of communication between neighboring GPUs doubled in the new simulation while the total amount of computation quadrupled.  Thus, communication is expected to have smaller impact on overall performance.  This is indeed the observed behavior, as scaling from 2 to 3 and 4 GPUs leads to improvements of $1.5\times$ and $2.2\times$ resp. in the larger-scale simulation, and only $1.27\times$ and $1.49\times$ in the smaller-scale experiment.
Thus, we expect that the scaling behavior can be further improved as we perform simulations at even larger scales.

\subsection{Comparison with GPU-based INSE solvers}

To further highlight the advantages in performance and accuracy of our GPU-optimized kinetic solver, we compared it with the recently proposed reflection-advection MacCormack (MC+R) solver~\cite{zehnder-2018}, which we implemented on the GPU using an SoA data layout and using NVIDIA's optimized cuSPARSE library~\cite{naumov2010cusparse} for solving sparse linear systems.
We note that a newer method called ``$\text{BiMocq}^2$''~\cite{qu-2019} can better preserve turbulence details and possibly offer enhanced performance compared to MC+R on the GPU.
However, there is no publicly available full GPU implementation of the $\text{BiMocq}^2$ method, making it difficult to compare against. In addition, based on the results from ~\cite{qu-2019} at similar grid resolutions, the method does not produce higher visual quality compared to our kinetic method. 
Thus, in the following performance analysis we base our comparison against the GPU-based MC+R INSE solver.


To ensure accuracy, especially for turbulent flow simulations, the preconditioned conjugate gradient (PCG) algorithm~\cite{helfenstein2012parallel} was used to solve the sparse linear systems using a sufficient number of iterations.
At a resolution of $200\!\times\!400\!\times\!200$, we found that at least 300 iterations were needed for proper simulation of turbulent flows when balancing accuracy and efficiency. 
The performance of INSE solvers can be improved by using larger time step sizes while still producing visually plausible results, especially in scenarios such as real time applications.  However, we found that in turbulent flow simulations, larger time step sizes sacrifice temporal accuracy and lead to results which deviate substantially from the reference output.  Below, we compare our simulation with those from the MC+R INSE solver at different time step sizes.

\begin{figure*}[t]
	\centering
	\includegraphics[width=\textwidth]{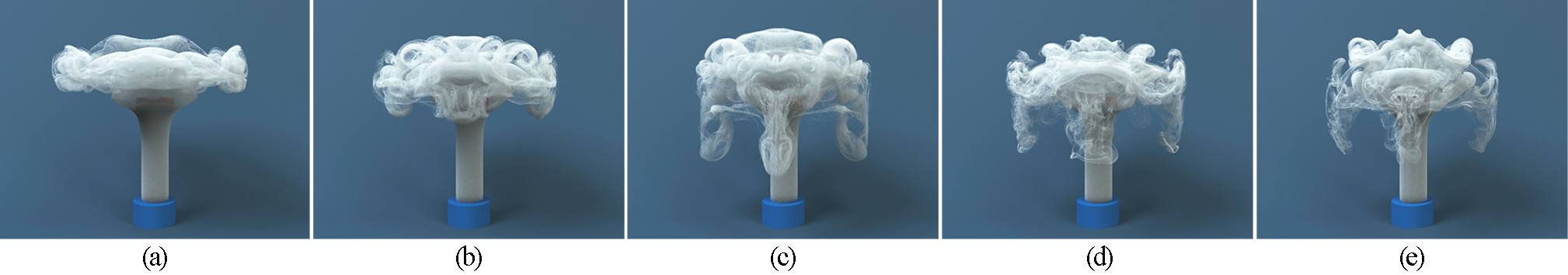}
	\vspace{-4mm}
	\caption{\textbf{Visual comparison of fluid solvers at different settings.} MC+R INSE solver with time step sizes (a) $40\times$,  (b) $10\times$ and (c) equal to our solver. (d) Output of our kinetic solver using the ACM-MRT model. (e) Output of our kinetic solver using a coarser grid resolution (half the resolution in each dimension). For the simulations in (a) to (d), the grid resolution is 200$\times$400$\times$200. All these simulations are implemented on GPUs using our optimizations. Note that (c) and (d) are visually similar (however, our kinetic solver can produce even finer details), while the INSE solver with larger time step sizes have much different smoke patterns due to temporal smoothing, smearing out turbulence details. The simulation in (a) has similar speed as ours, while (b) and (c) perform $5\times$ and $50\times$ slower respectively than our solver. In addition, due to its higher convergence rate, our kinetic solver operating at a lower grid resolution in (e) still produces similar smoke patterns to those shown in (d), but runs 6 times faster than (d) and 300 times faster than (c).  These results demonstrate the strength of our kinetic solver in both performance and visual quality.}
	\label{fig:ns-comparison-visual}
\end{figure*}

\begin{figure}[t]
	\centering
	\includegraphics[width=0.95\columnwidth]{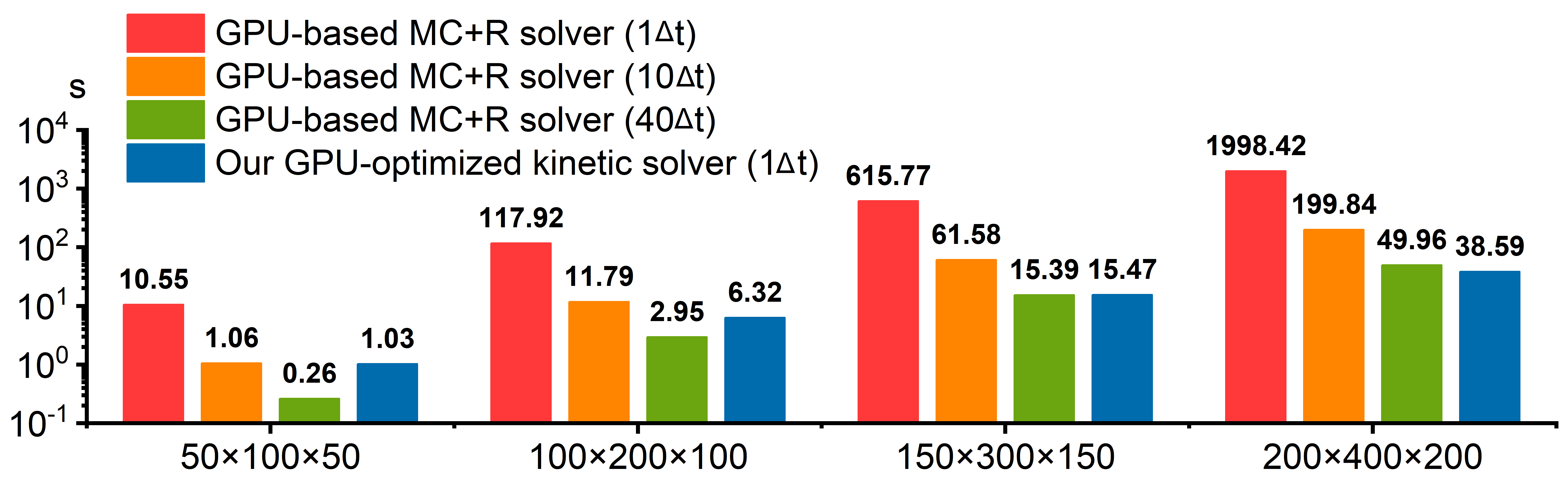}
	\vspace{-4mm}
	\caption{\textbf{Performance comparison with MC+R INSE solver using different time step sizes.} We show the timings (by simulating the same \textit{one physical second}) of a GPU-based MC+R INSE solver with varying time step sizes and our GPU-optimized kinetic solver under different grid resolutions. Due to large variations in timing, we plot the graph using a log-scale. Note the high scalability of our kinetic solver.}
	\label{fig:ns-comparison}
\end{figure}

Fig.~\ref{fig:ns-comparison-visual} shows a visual comparison between our kinetic solver and an MC+R INSE solver using different time step sizes.
Clearly, using a time step size which is (a) $40$ times (Fig.~\ref{fig:ns-comparison-visual} (a)) or (b) 10 times (Fig.~\ref{fig:ns-comparison-visual} (b)) larger than that used in our kinetic solver (Fig.~\ref{fig:ns-comparison-visual} (d)) can improve the INSE solver's performance to a degree that it becomes comparable to ours.  However, these step sizes also produce flow patterns which deviate substantially from the reference and lose large amounts of turbulence details.  The INSE simulation in Fig.~\ref{fig:ns-comparison-visual} (c) uses the same time step size as ours and produces similar flow patterns, which indicates the accuracy of our solver.  However, our solver has far superior performance, as demonstrated in Fig.~\ref{fig:ns-comparison} for different grid resolutions and with different time step sizes.

We note that the performance of INSE solvers can be further improved by techniques such as multi-grid methods.
Based on an existing report~\cite{Mcadams-2010}, multi-grid methods can typically accelerate INSE solvers by 10 to 20 times on a multi-core CPU, and can be expected to run even faster on a GPU.
However, a multi-grid INSE solver is still less scalable over multiple GPUs than our kinetic solver and thus less efficient when grid resolution is very high.  
In addition, our solver converges faster than many existing INSE solvers used in the graphics domain (as shown in e.g.~\cite{Li-2020}), implying that we can use lower grid resolutions while still producing visually similar results.  Fig.~\ref{fig:ns-comparison} (e) shows a simulation with our solver using only half the resolution in each dimension (and thus 8$\times$ fewer grid nodes) while producing patterns very similar to the higher resolution results in (c) and (d). Note that at the lower grid resolution, our solver is even faster than the MC+R INSE solver using a $40 \times$ larger time step size, demonstrating our solver can both run faster and produce higher quality results than state-of-the-art INSE solvers.  Lastly, our solver exhibits better scalability than GPU-based INSE solvers, enabling fast simulations at high grid resolutions.

\subsection{Visual simulation results}
\begin{figure}[t]
	\centering
	\includegraphics[width=\columnwidth]{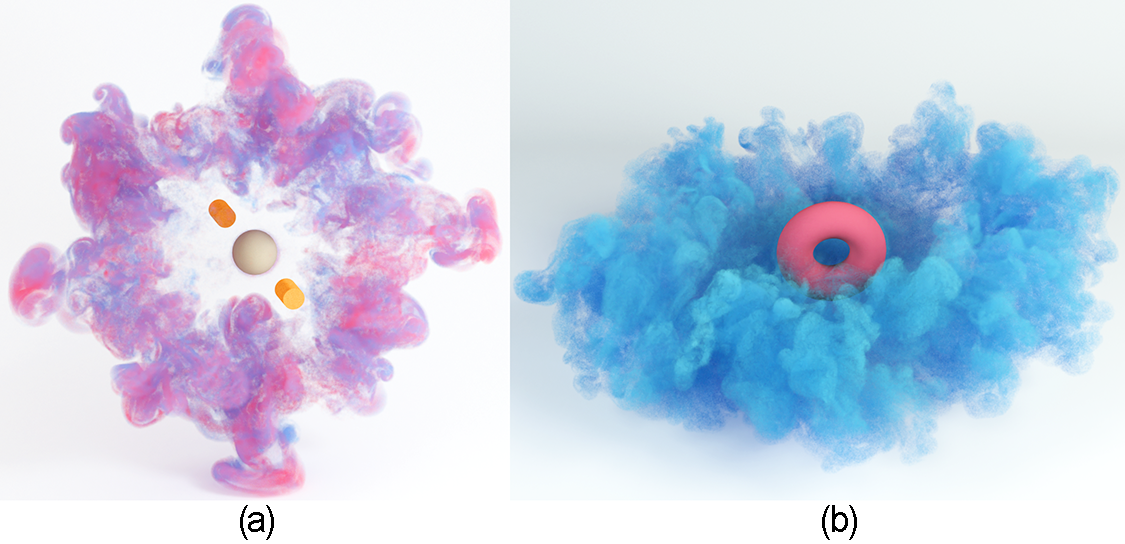}
	\vspace{-6mm}
	\caption{\textbf{Smoke simulations at a normal resolution}. (a) simulation of vortices colliding around a ball; (b) simulation of a torus rotating in the air. Both simulations use a grid resolution of 400$\times$200$\times$400.}
	\label{fig:smoke-collide}
\end{figure}

\begin{figure}[t]
	\centering
	\includegraphics[width=\columnwidth]{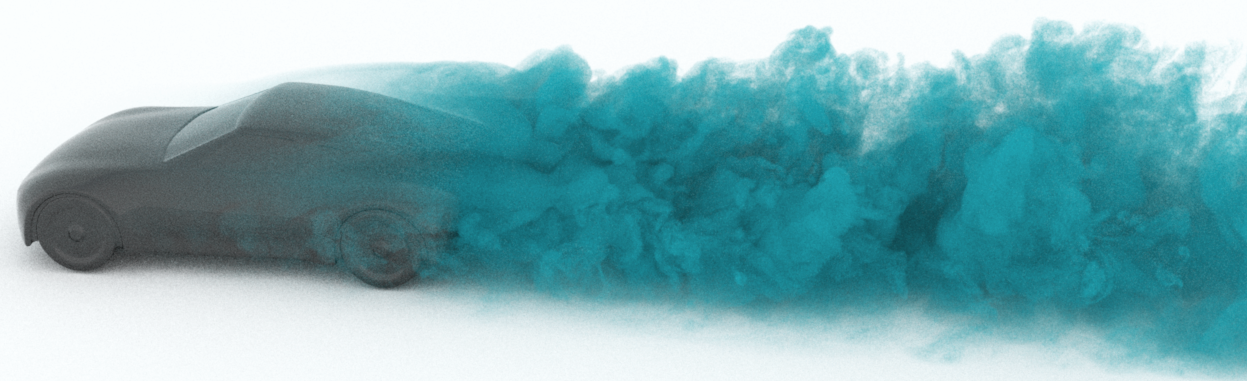} \vspace{-4mm}
	\caption{\textbf{High resolution simulation of airflow around a car}. We employed 4 GPUs to simulate the airflow over a car at a high resolution, displaying realistic wake turbulence details behind the car.}
	\label{fig:high-res-result-car}
\end{figure}

\begin{figure}[t]
	\centering
	\includegraphics[width=\columnwidth]{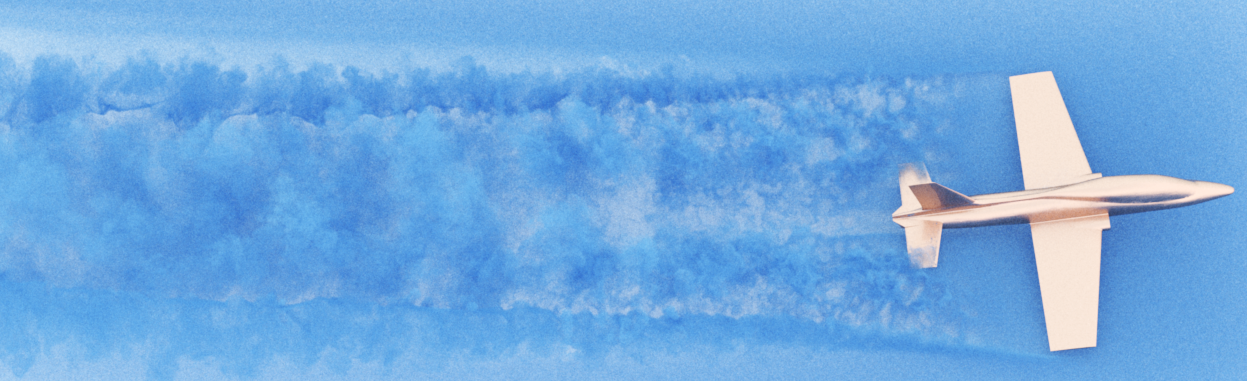} 
	\vspace{-4mm}
	\caption{\textbf{High resolution simulation of a moving airplane}. We employed 4 GPUs to simulate an airplane moving through high resolution smoke and generating detailed wake turbulence.}
	\label{fig:high-res-airplane}
\end{figure}

Using our high-performance kinetic solver, we can efficiently generate a variety of fluid simulation results.
Fig. \ref{fig:smoke-collide} shows two flow simulations of colliding vortices where a ball obstacle is immersed in the center and a rotating torus disturbs the surrounding smoke. To enable high resolution simulations, we ran our solver on a 4-GPU system.  Figs. \ref{fig:teaser}, \ref{fig:high-res-result-car} and \ref{fig:high-res-airplane} show high resolution simulations involving complex solid objects (both static and dynamic) computed on the multi-GPU system within a relatively short amount of time and displaying detailed and realistic-looking turbulence.


\subsection{Model parameter variation}
The optimal values of model parameters $\ell$ and $\alpha$ may change depending on the shape of the immersed solid as well as the number of solid samples within each fluid cell.
For simple geometries such as a ball or a torus, the cost for different $\ell$ and $\alpha$ values is similar, as shown in Fig.~\ref{fig:parameter-performance} (a).  The optimal $\ell$ lies in the range $[2,4]$, while the optimal $\alpha \geq 64 = 2^6$; picking any $\ell$ and $\alpha$ in this range leads to acceptable performance.  However, for complex geometries with high degrees of concavity, the cost surface exhibits significant variations with a number of local peaks, as shown in  Fig. \ref{fig:parameter-performance} (b).
If such a geometry also contains a large number of solid samples, the cost surface becomes even more unpredictable, as shown in Fig.~\ref{fig:parameter-performance} (c).
Thus, conducting an automated search for optimal parameter values before the start of a simulation is very useful in complex simulation scenarios.

\subsection{Limitations}
While our kinetic solver is in general much faster than INSE solvers and produces equal or superior visual results, one of its drawbacks is the need for larger amounts of memory, typically 3 times more than the MC+R INSE solver at the same grid resolution.  
In addition, in order to increase GPU occupancy, we employed multiple kernel launches, which further increases the memory requirement.
In situations where memory is limited, we can eliminate multiple kernel launches to save memory, but at the cost of lower occupancy and reduced simulation efficiency.
However, we note that even in these cases, our solver is still much faster than GPU-based INSE solvers.  Furthermore, for high resolution simulations with dynamic solids, we currently store all solid samples in each GPU to avoid communication, which again leads to a moderate increase in memory usage.  Finally, our GPU optimizations are currently not suitable for liquid simulations with kinetic solvers.  These simulations would need modifications to the kinetic algorithm, such as the use of a volume-of-fluid method, and thus require revising our memory layout.
These limitations point toward further research on more versatile and memory-efficient GPU-based kinetic solvers.

%% file: 7.conclusion.tex
\section{Conclusion}
In this paper, we proposed GPU optimization techniques to significantly improve the performance of an IB-based kinetic solver using the recently developed ACM-MRT model.  Our solver is faster than all other currently available fluid simulation methods, especially for turbulent flow simulations over complex solid geometries.  It also produces equal or superior visual quality compared to state-of-the-art INSE solvers.  Our optimizations are based on a parametric data layout which considers both fluid nodes and solid samples to improve memory coalescing and bandwidth use.  We also proposed a GPU-optimized parallel algorithm for the immersed boundary method to handle boundary conditions, which significantly reduces load imbalance and thread divergence.  We used multiple kernel launches with a simplified formulation of the ACM-MRT collision model to improve GPU occupancy.  Finally, we combined these optimizations into an integrated cost model to automatically search foroptimal parameter settings to obtain the best performance.   We conducted comprehensive comparisons against a range of state-of-the-art solution methods and solvers to validate the high efficiency and visual fidelity of our method.  In the future, we will further optimize our solver, including improving its performance in multi-GPU systems and scaling to larger problem sizes.

%% file: lbm-gpu-acc.bbl
\begin{thebibliography}{100}
\providecommand{\url}[1]{#1}
\csname url@samestyle\endcsname
\providecommand{\newblock}{\relax}
\providecommand{\bibinfo}[2]{#2}
\providecommand{\BIBentrySTDinterwordspacing}{\spaceskip=0pt\relax}
\providecommand{\BIBentryALTinterwordstretchfactor}{4}
\providecommand{\BIBentryALTinterwordspacing}{\spaceskip=\fontdimen2\font plus
\BIBentryALTinterwordstretchfactor\fontdimen3\font minus
  \fontdimen4\font\relax}
\providecommand{\BIBforeignlanguage}[2]{{%
\expandafter\ifx\csname l@#1\endcsname\relax
\typeout{** WARNING: IEEEtran.bst: No hyphenation pattern has been}%
\typeout{** loaded for the language `#1'. Using the pattern for}%
\typeout{** the default language instead.}%
\else
\language=\csname l@#1\endcsname
\fi
#2}}
\providecommand{\BIBdecl}{\relax}
\BIBdecl

\bibitem{Li-2018}
W.~{Li}, K.~{Bai}, and X.~{Liu}, ``Continuous-scale kinetic fluid simulation,''
  \emph{IEEE Transactions on Visualization and Computer Graphics}, vol.~25,
  no.~9, pp. 2694--2709, Sep. 2019.

\bibitem{Li-2020}
W.~Li, Y.~Chen, M.~Desbrun, C.~Zheng, and X.~Liu, ``Fast and scalable turbulent
  flow simulation with two-way coupling,'' \emph{ACM Transactions on Graphics
  (SIGGRAPH 2020)}, vol.~39, no.~4, 2020.

\bibitem{mullen2009energy}
P.~Mullen, K.~Crane, D.~Pavlov, Y.~Tong, and M.~Desbrun, ``Energy-preserving
  integrators for fluid animation,'' in \emph{ACM Transactions on Graphics
  (SIGGRAPH 2009)}, vol.~28, no.~3.\hskip 1em plus 0.5em minus 0.4em\relax ACM,
  2009, p.~38.

\bibitem{zhu2013new}
B.~Zhu, W.~Lu, M.~Cong, B.~Kim, and R.~Fedkiw, ``A new grid structure for
  domain extension,'' \emph{ACM Transactions on Graphics (SIGGRAPH 2013)},
  vol.~32, no.~4, p.~63, 2013.

\bibitem{ihmsen2014sph}
M.~Ihmsen, J.~Orthmann, B.~Solenthaler, A.~Kolb, and M.~Teschner, ``{SPH}
  fluids in computer graphics,'' 2014.

\bibitem{fu-2017}
C.~Fu, Q.~Guo, T.~Gast, C.~Jiang, and J.~Teran, ``A polynomial particle-in-cell
  method,'' \emph{ACM Transactions on Graphics}, vol.~36, no.~6, p. 222, 2017.

\bibitem{zehnder-2018}
J.~Zehnder, R.~Narain, and B.~Thomaszewski, ``An advection-reflection solver
  for detail-preserving fluid aimulation,'' \emph{ACM Transactions on Graphics
  (SIGGRAPH 2018)}, vol.~37, no.~4, p.~85, 2018.

\bibitem{qu-2019}
Z.~Qu, X.~Zhang, M.~Gao, C.~Jiang, and B.~Chen, ``Efficient and conservative
  fluids using bidirectional mapping,'' \emph{ACM Transactions on Graphics
  (SIGGRAPH 2019)}, vol.~38, no.~4, p. 128, 2019.

\bibitem{shan-2006}
X.~Shan, X.-F. Yuan, and H.~Chen, ``Kinetic theory representation of
  hydrodynamics: A way beyond the {Navier-Stokes} equation,'' \emph{Journal of
  Fluid Mechanics}, vol. 550, pp. 413--441, 2006.

\bibitem{Guo_TVCG_2017}
Y.~Guo, X.~Liu, and X.~Xu, ``A unified detail-preserving liquid simulation by
  two-phase lattice {Boltzmann} modeling,'' \emph{IEEE Transactions on
  Visualization and Computer Graphics}, vol.~23, no.~5, pp. 1479--1491, May
  2017.

\bibitem{wei-2004}
X.~Wei, Y.~Zhao, Z.~Fan, W.~Li, F.~Qiu, S.~Yoakum-Stover, and A.~E. Kaufman,
  ``Lattice-based flow field modeling,'' \emph{IEEE Transactions on
  Visualization and Computer Graphics}, vol.~10, no.~6, pp. 719--729, 2004.

\bibitem{Thuerey-2006}
N.~Thuerey, K.~Iglberger, and U.~Ruede, ``Free surface flows with moving and
  deforming objects for {LBM},'' \emph{Proceedings of Vision, Modeling and
  Visualization 2006}, pp. 193--200, Nov 2006.

\bibitem{Thuerey-2008}
N.~Thuerey and U.~Ruede, ``Stable free surface flows with the lattice
  {Boltzmann} method on adaptively coarsened grids,'' \emph{Computing and
  Visualization in Science}, vol. 12 (5), 2009.

\bibitem{chen1998lattice}
S.~Chen and G.~D. Doolen, ``Lattice {Boltzmann} method for fuid flows,''
  \emph{Annual review of fluid mechanics}, vol.~30, no.~1, pp. 329--364, 1998.

\bibitem{d2002multiple}
D.~d'Humieres, ``Multiple-relaxation-time lattice {Boltzmann} models in three
  dimensions,'' \emph{Philosophical Transactions of the Royal Society of
  London. Series A: Mathematical, Physical and Engineering Sciences}, vol. 360,
  no. 1792, pp. 437--451, 2002.

\bibitem{liu2012turbulence}
X.~Liu, W.-M. Pang, J.~Qin, and C.-W. Fu, ``Turbulence simulation by adaptive
  multi-relaxation lattice {Boltzmann} modeling,'' \emph{IEEE transactions on
  visualization and computer graphics}, vol.~20, no.~2, pp. 289--302, 2012.

\bibitem{Peskin-1972}
C.~S. Peskin, ``Flow patterns around heart valves: A numerical method,''
  \emph{Journal of computational physics}, vol.~10, no.~2, pp. 252--271, 1972.

\bibitem{herschlag2018gpu}
G.~Herschlag, S.~Lee, J.~S. Vetter, and A.~Randles, ``{GPU} data access on
  complex geometries for {D3Q19} lattice {Boltzmann} method,'' in \emph{2018
  IEEE International Parallel and Distributed Processing Symposium
  (IPDPS)}.\hskip 1em plus 0.5em minus 0.4em\relax IEEE, 2018, pp. 825--834.

\bibitem{calore2019optimization}
E.~Calore, A.~Gabbana, S.~F. Schifano, and R.~Tripiccione, ``Optimization of
  lattice {Boltzmann} simulations on heterogeneous computers,'' \emph{The
  International Journal of High Performance Computing Applications}, vol.~33,
  no.~1, pp. 124--139, 2019.

\bibitem{tolke2008teraflop}
J.~T{\"o}lke and M.~Krafczyk, ``Teraflop computing on a desktop {PC} with {GPUs
  for 3D CFD},'' \emph{International Journal of Computational Fluid Dynamics},
  vol.~22, no.~7, pp. 443--456, 2008.

\bibitem{Chen-1998}
S.~Chen and G.~D. Doolen, ``Lattice {Boltzmann} method for fluid flows,''
  \emph{Annual review of fluid mechanics}, vol.~30, no.~1, pp. 329--364, 1998.

\bibitem{stam1999stable}
J.~Stam, ``Stable fluids,'' in \emph{SIGGRAPH}, vol.~99, 1999, pp. 121--128.

\bibitem{kim2005flowfixer}
B.~Kim, Y.~Liu, I.~Llamas, and J.~R. Rossignac, ``Flowfixer: Using {BFECC} for
  fluid simulation,'' Georgia Institute of Technology, Tech. Rep., 2005.

\bibitem{wang2008observations}
R.~Wang, H.~Feng, and R.~J. Spiteri, ``Observations on the fifth-order {WENO}
  method with non-uniform meshes,'' \emph{Applied Mathematics and Computation},
  vol. 196, no.~1, pp. 433--447, 2008.

\bibitem{selle2008unconditionally}
A.~Selle, R.~Fedkiw, B.~Kim, Y.~Liu, and J.~Rossignac, ``An unconditionally
  stable {MacCormack} method,'' \emph{Journal of Scientific Computing},
  vol.~35, no. 2-3, pp. 350--371, 2008.

\bibitem{cui-2018}
Q.~Cui, P.~Sen, and T.~Kim, ``Scalable {Laplacian} eigenfluids,'' \emph{ACM
  Transactions on Graphics (SIGGRAPH 2018)}, vol.~37, no.~4, p.~87, 2018.

\bibitem{fedkiw2001visual}
R.~Fedkiw, J.~Stam, and H.~W. Jensen, ``Visual simulation of smoke,'' in
  \emph{Proceedings of the 28th annual conference on Computer graphics and
  interactive techniques (SIGGRAPH 2001)}.

\bibitem{park2005vortex}
S.~I. Park and M.~J. Kim, ``Vortex fluid for gaseous phenomena,'' in
  \emph{Proceedings of the 2005 ACM SIGGRAPH/Eurographics symposium on Computer
  animation}.\hskip 1em plus 0.5em minus 0.4em\relax ACM, 2005, pp. 261--270.

\bibitem{weissmann2010filament}
S.~Wei{\ss}mann and U.~Pinkall, ``Filament-based smoke with vortex shedding and
  variational reconnection,'' in \emph{ACM Transactions on Graphics (SIGGRAPH
  2010)}, vol.~29, no.~4.\hskip 1em plus 0.5em minus 0.4em\relax Citeseer,
  2010, p. 115.

\bibitem{pfaff2010scalable}
T.~Pfaff, N.~Thuerey, J.~Cohen, S.~Tariq, and M.~Gross, ``Scalable fluid
  simulation using anisotropic turbulence particles,'' in \emph{ACM
  Transactions on Graphics (SIGGRAPH Asia 2010)}, vol.~29, no.~6.\hskip 1em
  plus 0.5em minus 0.4em\relax ACM, 2010, p. 174.

\bibitem{brochu2012linear}
T.~Brochu, T.~Keeler, and R.~Bridson, ``Linear-time smoke animation with vortex
  sheet meshes,'' in \emph{Proceedings of the ACM SIGGRAPH/Eurographics
  Symposium on Computer Animation}.\hskip 1em plus 0.5em minus 0.4em\relax
  Eurographics Association, 2012, pp. 87--95.

\bibitem{golas2012large}
A.~Golas, R.~Narain, J.~Sewall, P.~Krajcevski, P.~Dubey, and M.~Lin,
  ``Large-scale fluid simulation using velocity-vorticity domain
  decomposition,'' \emph{ACM Transactions on Graphics (SIGGRAPH Asia 2012)},
  vol.~31, no.~6, p. 148, 2012.

\bibitem{zhang2014pppm}
X.~Zhang and R.~Bridson, ``A {PPPM} fast summation method for fluids and
  beyond,'' \emph{ACM Transactions on Graphics (SIGGRAPH 2014)}, vol.~33,
  no.~6, p. 206, 2014.

\bibitem{zhang-2015}
X.~Zhang, R.~Bridson, and C.~Greif, ``Restoring the missing vorticity in
  advection-projection fluid solvers,'' \emph{ACM Transactions on Graphics
  (SIGGRAPH 2015)}, vol.~34, no.~4, pp. 52:1--52:8, Jul. 2015.

\bibitem{bridson2007curl}
R.~Bridson, J.~Houriham, and M.~Nordenstam, ``Curl-noise for procedural fluid
  flow,'' in \emph{ACM Transactions on Graphics (SIGGRAPH 2007)}, vol.~26,
  no.~3.\hskip 1em plus 0.5em minus 0.4em\relax ACM, 2007, p.~46.

\bibitem{kim2008wavelet}
T.~Kim, N.~Th{\"u}erey, D.~James, and M.~Gross, ``Wavelet turbulence for fluid
  simulation,'' in \emph{ACM Transactions on Graphics (SIGGRAPH 2008)},
  vol.~27, no.~3.\hskip 1em plus 0.5em minus 0.4em\relax ACM, 2008, p.~50.

\bibitem{jeong2015data}
S.~Jeong, B.~Solenthaler, M.~Pollefeys, M.~Gross \emph{et~al.}, ``Data-driven
  fluid simulations using regression forests,'' \emph{ACM Transactions on
  Graphics (SIGGRAPH 2015)}, vol.~34, no.~6, p. 199, 2015.

\bibitem{chu2017data}
M.~Chu and N.~Thuerey, ``Data-driven synthesis of smoke flows with {CNN-based}
  feature descriptors,'' \emph{ACM Transactions on Graphics (SIGGRAPH 2017)},
  vol.~36, no.~4, p.~69, 2017.

\bibitem{xie-2018}
Y.~Xie, E.~Franz, M.~Chu, and N.~Thuerey, ``{TempoGAN}: A temporally coherent,
  volumetric {GAN} for super-resolution fluid flow,'' \emph{ACM Transactions on
  Graphics (SIGGRAPH 2018)}, vol.~37, no.~4, p.~95, 2018.

\bibitem{Bai-2020}
\BIBentryALTinterwordspacing
K.~Bai, W.~Li, M.~Desbrun, and X.~Liu, ``Dynamic upsampling of smoke through
  dictionary-based learning,'' \emph{ACM Transactions on Graphics}, vol.~40,
  no.~1, Sep. 2020. [Online]. Available: \url{https://doi.org/10.1145/3412360}
\BIBentrySTDinterwordspacing

\bibitem{losasso2004simulating}
F.~Losasso, F.~Gibou, and R.~Fedkiw, ``Simulating water and smoke with an
  octree data structure,'' in \emph{ACM Transactions on Graphics (SIGGRAPH
  2004)}, vol.~23, no.~3.\hskip 1em plus 0.5em minus 0.4em\relax ACM, 2004, pp.
  457--462.

\bibitem{setaluri2014spgrid}
R.~Setaluri, M.~Aanjaneya, S.~Bauer, and E.~Sifakis, ``{SPGrid}: A sparse paged
  grid structure applied to adaptive smoke simulation,'' \emph{ACM Transactions
  on Graphics (SIGGRAPH 2014)}, vol.~33, no.~6, p. 205, 2014.

\bibitem{clausen-2013}
P.~Clausen, M.~Wicke, J.~R. Shewchuk, and J.~F. O'brien, ``Simulating liquids
  and solid-liquid interactions with {Lagrangian} meshes,'' \emph{ACM
  Transactions on Graphics}, vol.~32, no.~2, p.~17, 2013.

\bibitem{ando2013highly}
R.~Ando, N.~Th{\"u}rey, and C.~Wojtan, ``Highly adaptive liquid simulations on
  tetrahedral meshes,'' \emph{ACM Transactions on Graphics (SIGGRAPH 2013)},
  vol.~32, no.~4, p. 103, 2013.

\bibitem{becker2007weakly}
M.~Becker and M.~Teschner, ``Weakly compressible {SPH} for free surface
  flows,'' in \emph{Proceedings of the 2007 ACM SIGGRAPH/Eurographics symposium
  on Computer animation}.\hskip 1em plus 0.5em minus 0.4em\relax Eurographics
  Association, 2007, pp. 209--217.

\bibitem{solenthaler2009predictive}
B.~Solenthaler and R.~Pajarola, ``Predictive-corrective incompressible {SPH},''
  in \emph{ACM Transactions on Graphics (SIGGRAPH Asia 2009)}, vol.~28,
  no.~3.\hskip 1em plus 0.5em minus 0.4em\relax ACM, 2009, p.~40.

\bibitem{akinci2012versatile}
N.~Akinci, M.~Ihmsen, G.~Akinci, B.~Solenthaler, and M.~Teschner, ``Versatile
  rigid-fluid coupling for incompressible {SPH},'' \emph{ACM Transactions on
  Graphics (SIGGRAPH Asia 2009)}, vol.~31, no.~4, p.~62, 2012.

\bibitem{winchenbach2017infinite}
R.~Winchenbach, H.~Hochstetter, and A.~Kolb, ``Infinite continuous adaptivity
  for incompressible {SPH},'' \emph{ACM Transactions on Graphics (SIGGRAPH
  2017)}, vol.~36, no.~4, p. 102, 2017.

\bibitem{de2015power}
F.~de~Goes, C.~Wallez, J.~Huang, D.~Pavlov, and M.~Desbrun, ``Power particles:
  An incompressible fluid solver based on power diagrams,'' \emph{ACM
  Transactions on Graphics (SIGGRAPH 2015)}, vol.~34, no.~4, pp. 50--1, 2015.

\bibitem{bender-2016}
J.~Bender and D.~Koschier, ``Divergence-free {SPH} for incompressible and
  viscous fluids,'' \emph{IEEE Transactions on Visualization and Computer
  Graphics}, vol.~23, no.~3, pp. 1193--1206, 2016.

\bibitem{band-2017}
S.~Band, C.~Gissler, and M.~Teschner, ``Moving least squares boundaries for
  {SPH} fluids,'' in \emph{Proceedings of the 13th Workshop on Virtual Reality
  Interactions and Physical Simulations}.\hskip 1em plus 0.5em minus
  0.4em\relax Eurographics Association, 2017, pp. 21--28.

\bibitem{Band-2018}
\BIBentryALTinterwordspacing
S.~Band, C.~Gissler, M.~Ihmsen, J.~Cornelis, A.~Peer, and M.~Teschner,
  ``Pressure boundaries for implicit incompressible {SPH},'' \emph{ACM
  Transactions on Graphics (SIGGRAPH 2018)}, vol.~37, no.~2, pp. 14:1--14:11,
  Feb. 2018. [Online]. Available: \url{http://doi.acm.org/10.1145/3180486}
\BIBentrySTDinterwordspacing

\bibitem{jiang2015affine}
C.~Jiang, C.~Schroeder, A.~Selle, J.~Teran, and A.~Stomakhin, ``The affine
  particle-in-cell method,'' \emph{ACM Transactions on Graphics (SIGGRAPH
  2015)}, vol.~34, no.~4, p.~51, 2015.

\bibitem{zhang2016resolving}
X.~Zhang, M.~Li, and R.~Bridson, ``Resolving fluid boundary layers with
  particle strength exchange and weak adaptivity,'' \emph{ACM Transactions on
  Graphics (SIGGRAPH 2016)}, vol.~35, no.~4, p.~76, 2016.

\bibitem{zhu2005animating}
Y.~Zhu and R.~Bridson, ``Animating sand as a fluid,'' in \emph{ACM Transactions
  on Graphics (SIGGRAPH 2005)}, vol.~24, no.~3.\hskip 1em plus 0.5em minus
  0.4em\relax ACM, 2005, pp. 965--972.

\bibitem{geier2006cascaded}
M.~Geier, A.~Greiner, and J.~G. Korvink, ``Cascaded digital lattice {Boltzmann}
  automata for high reynolds number flow,'' \emph{Physical Review E}, vol.~73,
  no.~6, p. 066705, 2006.

\bibitem{lycett2014binary}
D.~Lycett-Brown, K.~H. Luo, R.~Liu, and P.~Lv, ``Binary droplet collision
  simulations by a multiphase cascaded lattice {Boltzmann} method,''
  \emph{Physics of Fluids}, vol.~26, no.~2, p. 023303, 2014.

\bibitem{geier-2009}
M.~Geier, A.~Greiner, and J.~G. Korvink, ``A factorized central moment lattice
  {Boltzmann} method,'' \emph{The European Physical Journal Special Topics},
  vol. 171, no.~1, pp. 55--61, 2009.

\bibitem{shan-2019}
X.~Shan \emph{et~al.}, ``Central-moment-based {Galilean}-invariant
  multiple-relaxation-time collision model,'' \emph{Physical Review E}, vol.
  100, no.~4, p. 043308, 2019.

\bibitem{de2017nonorthogonal}
A.~De~Rosis, ``Nonorthogonal central-moments-based lattice {Boltzmann} scheme
  in three dimensions,'' \emph{Physical Review E}, vol.~95, no.~1, p. 013310,
  2017.

\bibitem{rohde-2003}
M.~Rohde, D.~Kandhai, J.~Derksen, and H.~Van~den Akker, ``Improved bounce-back
  methods for no-slip walls in lattice-{Boltzmann} schemes: Theory and
  simulations,'' \emph{Physical Review E}, vol.~67, no.~6, p. 066703, 2003.

\bibitem{sbragaglia-2005}
M.~Sbragaglia and S.~Succi, ``Analytical calculation of slip flow in lattice
  {Boltzmann} models with kinetic boundary conditions,'' \emph{Physics of
  Fluids}, vol.~17, no.~9, p. 093602, 2005.

\bibitem{verschaeve-2010}
J.~C. Verschaeve and B.~M{\"u}ller, ``A curved no-slip boundary condition for
  the lattice {Boltzmann} method,'' \emph{Journal of Computational Physics},
  vol. 229, no.~19, pp. 6781--6803, 2010.

\bibitem{mei-1999}
R.~Mei, L.-S. Luo, and W.~Shyy, ``An accurate curved boundary treatment in the
  lattice {Boltzmann} method,'' \emph{Journal of computational physics}, vol.
  155, no.~2, pp. 307--330, 1999.

\bibitem{feng-2004}
Z.-G. Feng and E.~E. Michaelides, ``The immersed boundary-lattice {Boltzmann}
  method for solving fluid-particles interaction problems,'' \emph{Journal of
  Computational Physics}, vol. 195, no.~2, pp. 602--628, 2004.

\bibitem{wu-2010}
J.~Wu and C.~Shu, ``An improved immersed boundary-lattice {Boltzmann} method
  for simulating three-dimensional incompressible flows,'' \emph{Journal of
  Computational Physics}, vol. 229, no.~13, pp. 5022--5042, 2010.

\bibitem{sato-2013}
Y.~Sato, T.~Hino, and K.~Ohashi, ``Parallelization of an unstructured
  {Navier-Stokes} solver using a multi-color ordering method for openmp,''
  \emph{Computers \& Fluids}, vol.~88, pp. 496--509, 2013.

\bibitem{guo-2015}
X.~Guo, M.~Lange, G.~Gorman, L.~Mitchell, and M.~Weiland, ``Developing a
  scalable hybrid {MPI/OpenMP} unstructured finite element model,''
  \emph{Computers \& Fluids}, vol. 110, pp. 227--234, 2015.

\bibitem{mashayekhi-2018}
O.~Mashayekhi, C.~Shah, H.~Qu, A.~Lim, and P.~Levis, ``Automatically
  distributing {Eulerian} and hybrid fluid simulations in the cloud,''
  \emph{ACM Transactions on Graphics (SIGGRAPH 2018)}, vol.~37, no.~2, p.~24,
  2018.

\bibitem{brandvik-2008}
T.~Brandvik and G.~Pullan, ``Acceleration of a {3D} {Euler} solver using
  commodity graphics hardware,'' in \emph{46th AIAA aerospace sciences meeting
  and exhibit}, 2008, p. 607.

\bibitem{liu-2004}
Y.~Liu, X.~Liu, and E.~Wu, ``Real-time {3D} fluid simulation on {GPU} with
  complex obstacles,'' in \emph{12th Pacific Conference on Computer Graphics
  and Applications, 2004. PG 2004. Proceedings.}\hskip 1em plus 0.5em minus
  0.4em\relax IEEE, 2004, pp. 247--256.

\bibitem{kruger-2003}
J.~Kr{\"u}ger and R.~Westermann, ``Linear algebra operators for {GPU}
  implementation of numerical algorithms,'' in \emph{ACM Transactions on
  Graphics (SIGGRAPH 2003)}, vol.~22, no.~3.\hskip 1em plus 0.5em minus
  0.4em\relax ACM, 2003, pp. 908--916.

\bibitem{Cuda}
\BIBentryALTinterwordspacing
``{CUDA},'' parallel Computing Platform. [Online]. Available:
  \url{https://developer.nvidia.com/cuda-zone}
\BIBentrySTDinterwordspacing

\bibitem{cusparse}
\BIBentryALTinterwordspacing
``cusparse,'' sparse Linear Algebra. [Online]. Available:
  \url{https://developer.nvidia.com/cusparse}
\BIBentrySTDinterwordspacing

\bibitem{thibault-2009}
J.~Thibault and I.~Senocak, ``{CUDA} implementation of a {Navier-Stokes} solver
  on multi-{GPU} desktop platforms for incompressible flows,'' in \emph{47th
  AIAA aerospace sciences meeting including the new horizons forum and
  aerospace exposition}, 2009, p. 758.

\bibitem{griebel-2010}
M.~Griebel and P.~Zaspel, ``A multi-{GPU} accelerated solver for the
  three-dimensional two-phase incompressible {Navier-Stokes} equations,''
  \emph{Computer Science-Research and Development}, vol.~25, no. 1-2, pp.
  65--73, 2010.

\bibitem{henniger-2010}
R.~Henniger, D.~Obrist, and L.~Kleiser, ``High-order accurate solution of the
  incompressible {Navier-Stokes} equations on massively parallel computers,''
  \emph{Journal of Computational Physics}, vol. 229, no.~10, pp. 3543--3572,
  2010.

\bibitem{Mcadams-2010}
A.~McAdams, E.~Sifakis, and J.~Teran, ``A parallel multigrid poisson solver for
  fluids simulation on large grids,'' ser. SCA '10.\hskip 1em plus 0.5em minus
  0.4em\relax Goslar, DEU: Eurographics Association, 2010, p. 65–74.

\bibitem{gao-2018}
M.~Gao, X.~Wang, K.~Wu, A.~Pradhana, E.~Sifakis, C.~Yuksel, and C.~Jiang,
  ``{GPU} optimization of material point methods,'' in \emph{ACM Transactions
  on Graphics (SIGGRAPH Asia 2018)}.\hskip 1em plus 0.5em minus 0.4em\relax
  ACM, 2018, p. 254.

\bibitem{wu-2018}
K.~Wu, N.~Truong, C.~Yuksel, and R.~Hoetzlein, ``Fast fluid simulations with
  sparse volumes on the {GPU},'' in \emph{Computer Graphics Forum}, vol.~37,
  no.~2.\hskip 1em plus 0.5em minus 0.4em\relax Wiley Online Library, 2018, pp.
  157--167.

\bibitem{hu-2019}
Y.~Hu, T.-M. Li, L.~Anderson, J.~Ragan-Kelley, and F.~Durand, ``Taichi: A
  language for high-performance computation on spatially sparse data
  structures,'' \emph{ACM Transactions on Graphics (SIGGRAPH 2019)}, vol.~38,
  no.~6, pp. 1--16, 2019.

\bibitem{alfonsi-2011}
G.~Alfonsi, S.~A. Ciliberti, M.~Mancini, and L.~Primavera, ``Performances of
  {Navier-Stokes} solver on a hybrid {CPU/GPU} computing system,'' in
  \emph{International Conference on Parallel Computing Technologies}.\hskip 1em
  plus 0.5em minus 0.4em\relax Springer, 2011, pp. 404--416.

\bibitem{wang-2014}
Y.~Wang, M.~Baboulin, K.~Rupp, O.~Le~Ma{\^\i}tre, and Y.~Fraigneau, ``Solving
  {3D} incompressible {Navier-Stokes} equations on hybrid {CPU/GPU} systems,''
  in \emph{Proceedings of the High Performance Computing Symposium}.\hskip 1em
  plus 0.5em minus 0.4em\relax Society for Computer Simulation International,
  2014, p.~12.

\bibitem{posey-2013}
S.~Posey, ``Considerations for {GPU} acceleration of parallel {CFD},''
  \emph{Procedia Engineering}, vol.~61, pp. 388--391, 2013.

\bibitem{li2003implementing}
W.~Li, X.~Wei, and A.~Kaufman, ``Implementing lattice {Boltzmann} computation
  on graphics hardware,'' \emph{The Visual Computer}, vol.~19, no. 7-8, pp.
  444--456, 2003.

\bibitem{zhao2007flow}
Y.~Zhao, F.~Qiu, Z.~Fan, and A.~Kaufman, ``Flow simulation with locally-refined
  {LBM},'' in \emph{Proceedings of the 2007 Symposium on Interactive {3D}
  Graphics and Games}.\hskip 1em plus 0.5em minus 0.4em\relax ACM, 2007, pp.
  181--188.

\bibitem{WelleinOn2006}
G.~Wellein, T.~Zeiser, G.~Hager, and S.~Donath, ``On the single processor
  performance of simple lattice {Boltzmann} kernels,'' \emph{Computers \&
  Fluids}, vol.~35, no. 8-9, pp. 910--919.

\bibitem{MattilaAn2007}
K.~Mattila, J.~Hyv{\"a}luoma, T.~Rossi, M.~Aspn\"{a}s, and J.~Westerholm, ``An
  efficient swap algorithm for the lattice {B}oltzmann method,'' vol. 176,
  no.~3, pp. 200--210.

\bibitem{delbosc2014optimized}
N.~Delbosc, J.~L. Summers, A.~Khan, N.~Kapur, and C.~J. Noakes, ``Optimized
  implementation of the lattice {Boltzmann} method on a graphics processing
  unit towards real-time fluid simulation,'' \emph{Computers \& Mathematics
  with Applications}, vol.~67, no.~2, pp. 462--475, 2014.

\bibitem{tolke2010implementation}
J.~T{\"o}lke, ``Implementation of a lattice {Boltzmann} kernel using the
  compute unified device architecture developed by {NVIDIA},'' \emph{Computing
  and Visualization in Science}, vol.~13, no.~1, p.~29, 2010.

\bibitem{bailey2009accelerating}
P.~Bailey, J.~Myre, S.~D. Walsh, D.~J. Lilja, and M.~O. Saar, ``Accelerating
  lattice {Boltzmann} fluid flow simulations using graphics processors,'' in
  \emph{2009 international conference on parallel processing}.\hskip 1em plus
  0.5em minus 0.4em\relax IEEE, 2009, pp. 550--557.

\bibitem{habich2011performance}
J.~Habich, T.~Zeiser, G.~Hager, and G.~Wellein, ``Performance analysis and
  optimization strategies for a {D3Q19} lattice {Boltzmann} kernel on {NVIDIA
  GPUs Using CUDA},'' \emph{Advances in Engineering Software}, vol.~42, no.~5,
  pp. 266--272, 2011.

\bibitem{bernaschi2010flexible}
M.~Bernaschi, M.~Fatica, S.~Melchionna, S.~Succi, and E.~Kaxiras, ``A flexible
  high-performance lattice {Boltzmann GPU} code for the simulations of fluid
  flows in complex geometries,'' \emph{Concurrency and Computation: Practice
  and Experience}, vol.~22, no.~1, pp. 1--14, 2010.

\bibitem{obrecht2013multi}
C.~Obrecht, F.~Kuznik, B.~Tourancheau, and J.-J. Roux, ``{Multi-GPU}
  implementation of the lattice {Boltzmann} method,'' \emph{Computers \&
  Mathematics with Applications}, vol.~65, no.~2, pp. 252--261, 2013.

\bibitem{myre2011performance}
J.~Myre, S.~D. Walsh, D.~Lilja, and M.~O. Saar, ``Performance analysis of
  single-phase, multiphase, and multicomponent lattice-{Boltzmann} fluid flow
  simulations on {GPU} clusters,'' \emph{Concurrency and Computation: Practice
  and Experience}, vol.~23, no.~4, pp. 332--350, 2011.

\bibitem{harris-2004}
S.~Harris, \emph{An Introduction to the Theory of the {Boltzmann}
  Equation}.\hskip 1em plus 0.5em minus 0.4em\relax Courier Corporation, 2004.

\bibitem{girimaji2012lattice}
S.~Girimaji, ``Lattice {Boltzmann} method: Fundamentals and engineering
  applications with computer codes,'' 2012.

\bibitem{yuksel-2015}
C.~Yuksel, ``Sample elimination for generating {Poisson} disk sample sets,'' in
  \emph{Computer Graphics Forum}, vol.~34, no.~2.\hskip 1em plus 0.5em minus
  0.4em\relax Wiley Online Library, 2015, pp. 25--32.

\bibitem{kruger-2011}
T.~Kr{\"u}ger, F.~Varnik, and D.~Raabe, ``Efficient and accurate simulations of
  deformable particles immersed in a fluid using a combined immersed boundary
  lattice {Boltzmann} finite element method,'' \emph{Computers \& Mathematics
  with Applications}, vol.~61, no.~12, pp. 3485--3505, 2011.

\bibitem{mattila2008comparison}
K.~Mattila, J.~Hyv{\"a}luoma, J.~Timonen, and T.~Rossi, ``Comparison of
  implementations of the lattice-{Boltzmann} method,'' \emph{Computers \&
  Mathematics with Applications}, vol.~55, no.~7, pp. 1514--1524, 2008.

\bibitem{valero2015accelerating}
P.~Valero-Lara, F.~D. Igual, M.~Prieto-Mat{\'\i}as, A.~Pinelli, and J.~Favier,
  ``Accelerating fluid-solid simulations (lattice-{Boltzmann} \&
  immersed-boundary) on heterogeneous architectures,'' \emph{Journal of
  Computational Science}, vol.~10, pp. 249--261, 2015.

\bibitem{morton-1966}
G.~M. Morton, ``A computer oriented geodetic data base and a new technique in
  file sequencing,'' 1966.

\bibitem{heron1989particle}
A.~Heron and J.~Adam, ``Particle code optimization on vector computers,''
  \emph{Journal of Computational Physics}, vol.~85, no.~2, pp. 284--301, 1989.

\bibitem{JAKOB-2010}
\BIBentryALTinterwordspacing
Jakob, ``Mitsuba renderer,'' 2010. [Online]. Available:
  \url{http://www.mitsuba-renderer.org/}
\BIBentrySTDinterwordspacing

\bibitem{suga2015d3q27}
K.~Suga, Y.~Kuwata, K.~Takashima, and R.~Chikasue, ``A {D3Q27}
  multiple-relaxation-time lattice {Boltzmann} method for turbulent flows,''
  \emph{Computers \& Mathematics with Applications}, vol.~69, no.~6, pp.
  518--529, 2015.

\bibitem{naumov2010cusparse}
M.~Naumov, L.~Chien, P.~Vandermersch, and U.~Kapasi, ``{cuSPARSE} {Library},''
  in \emph{{GPU} Technology Conference}, 2010.

\bibitem{helfenstein2012parallel}
R.~Helfenstein and J.~Koko, ``Parallel preconditioned conjugate gradient
  algorithm on {GPU},'' \emph{Journal of Computational and Applied
  Mathematics}, vol. 236, no.~15, pp. 3584--3590, 2012.

\end{thebibliography}
